\documentclass[11pt]{article}

\usepackage[dvipsnames]{xcolor}
\usepackage{setspace}
\usepackage{comment}
\usepackage{atmacros}

\usepackage{fullpage,float}
\usepackage[T1]{fontenc}
\usepackage[shortlabels]{enumitem}
\usepackage{stmaryrd}
\usepackage{mathtools}

\usepackage{mathpazo}
\usepackage{bm}
\usepackage{todonotes}
\usepackage{lipsum}
\usepackage{scrextend}
\usepackage{xfrac}

\usepackage[linesnumbered,vlined,boxed,ruled]{algorithm2e}
\makeatletter
\newcommand{\RemoveAlgoNumber}{\renewcommand{\fnum@algocf}{\AlCapSty{\AlCapFnt\algorithmcfname}}}
\newcommand{\RevertAlgoNumber}{\algocf@resetfnum}
\makeatother

\RequirePackage[colorlinks=true, backref=page]{hyperref}
\hypersetup{
  linkcolor=[rgb]{0.3,0.3,0.6},
  citecolor=[rgb]{0.2, 0.6, 0.2},
  urlcolor=[rgb]{0.6, 0.2, 0.2}
}

\newcommand{\idperm}{\mathsf{id}}
\newcommand{\rev}{\operatorname{rev}}
\newcommand{\cutwidth}{\mathtt{CutWidth}}
\newcommand{\nisan}[3]{\mathrm{rank}\inparen{M^{#1}(#3)}}

\newcommand{\DenseROwidth}[1]{\mathtt{DenseROwidth}\text{-}{#1}}
\newcommand{\CktROwidth}{\mathtt{CktROwidth}}

\newcommand{\SearchCktROwidth}{\mathtt{Search}\text{-}\mathtt{CktROwidth}}
\newcommand{\SUBEXP}{\mathsf{SUBEXP}}
\newcommand{\inparen}[1]{\left( #1 \right)}
\newcommand{\insquare}[1]{\left[ #1 \right]}
\newcommand{\inbrace}[1]{\left\{ #1 \right\}}

\newcommand{\inangle}[1]{\left< #1 \right>}

\newcommand{\set}[1]{\inbrace{#1}}

\newcommand{\abs}[1]{\left| #1 \right|}
\newcommand{\floor}[1]{\left\lfloor #1 \right\rfloor}
\newcommand{\ceil}[1]{\left\lceil #1 \right\rceil}

\newcommand{\coeff}{\operatorname{coeff}}
\newcommand{\size}{\operatorname{size}}

\newcommand{\cvector}{\overline{\operatorname{coeff}}}

\def\P{\mathsf{P}}
\newcommand{\NP}{\mathsf{NP}}

\newcommand{\C}{\mathbb{C}}

\def\phi{\varphi}   
\def\epsilon{\varepsilon}   
\usepackage{amsthm}
\usepackage{thmtools,thm-restate}

\numberwithin{equation}{section}
\declaretheoremstyle[bodyfont=\it,qed=\qedsymbol]{noproofstyle}

\declaretheorem[numberlike=equation]{observation}
\declaretheorem[numberlike=equation,style=noproofstyle,name=Observation]{observationwp}
\declaretheorem[name=Observation,numbered=no]{observation*}

\declaretheorem[numberlike=equation]{fact}

\declaretheorem[numberlike=equation]{theorem}
\declaretheorem[numberlike=equation,style=noproofstyle,name=Theorem]{theoremwp}
\declaretheorem[name=Theorem,numbered=no]{theorem*}

\declaretheorem[numberlike=equation]{lemma}
\declaretheorem[name=Lemma,numbered=no]{lemma*}
\declaretheorem[numberlike=equation,style=noproofstyle,name=Lemma]{lemmawp}

\declaretheorem[numberlike=equation]{corollary}
\declaretheorem[name=Corollary,numbered=no]{corollary*}

\declaretheorem[name=Proposition,numbered=no]{proposition*}

\declaretheorem[numberlike=equation]{claim}
\declaretheorem[name=Claim,numbered=no]{claim*}

\declaretheorem[name=Conjecture,numbered=no]{conjecture*}

\declaretheorem[numberlike=equation]{question}
\declaretheorem[name=Question,numbered=no]{question*}

\declaretheoremstyle[bodyfont=\it,qed=$\lozenge$]{defstyle} 

\declaretheorem[numberlike=equation,style=defstyle]{definition}
\declaretheorem[unnumbered,name=Definition,style=defstyle]{definition*}

\declaretheorem[numberlike=equation,style=defstyle]{example}
\declaretheorem[unnumbered,name=Example,style=defstyle]{example*}

\declaretheorem[unnumbered,name=Notation=defstyle]{notation*}

\declaretheorem[unnumbered,name=Construction,style=defstyle]{construction*}

\declaretheorem[numberlike=equation,style=defstyle]{remark}
\declaretheorem[unnumbered,name=Remark,style=defstyle]{remark*}

\newcommand{\ehref}[1]{\href{mailto:#1}{#1}}
\newcommand{\ignore}[1]{}

\usepackage{nth}
\usepackage{intcalc}
\usepackage{etoolbox}
\usepackage{xstring}

\usepackage{ifpdf}
\ifpdf
\else
\usepackage[quadpoints=false]{hypdvips}
\fi

\newcommand{\ECCCurl}[2]{http://eccc.hpi-web.de/report/\ifnumcomp{#1}{>}{93}{19}{20}#1/#2/}

\newcommand{\shortECCC}[2]{\texttt{\href{http://eccc.hpi-web.de/report/\ifnumcomp{#1}{>}{93}{19}{20}#1/#2/}{eccc:TR#1-#2}}}

\newcommand{\parseECCC}[1]{
\StrSubstitute{#1}{TR}{}[\tmpstring]%
\IfSubStr{\tmpstring}{/}{ 
\StrBefore{\tmpstring}{/}[\ecccyear]%
\StrBehind{\tmpstring}{/}[\ecccreport]%
}{
\StrBefore{\tmpstring}{-}[\ecccyear]%
\StrBehind{\tmpstring}{-}[\ecccreport]%
}%
\shortECCC{\ecccyear}{\ecccreport}}

\newcommand{\ATnote}[1]{\textcolor{Magenta}{AT:#1}}

\title{The Complexity of Order-Finding for ROABPs}
\author{Vishwas Bhargava\thanks{\ehref{vishwas1384@gmail.com}. Department of Computing and Mathematical Sciences, Caltech. Part of this work was done as a postdoc at University of Waterloo, Canada.}%
\and%
Pranjal Dutta\thanks{\ehref{duttpranjal@gmail.com}. School of Computing, NUS.}
\and%
Sumanta Ghosh\thanks{\ehref{besusumanta@gmail.com}. Chennai Mathematical Institute.}
\and%
Anamay Tengse\thanks{\ehref{anamay.tengse@gmail.com}. School of Computer Sciences, NISER, Bhubaneswar. Initial parts of this work were done when the author was a postdoc at University of Haifa (supported by ISF grant no 716/20) and Reichman University, Herzliya (supported by ISF grant no 843/23).}}
\date{}

\begin{document}

\maketitle

\pagenumbering{gobble}

\begin{abstract}

We study the \emph{order-finding problem} for Read-once Oblivious Algebraic Branching Programs (ROABPs). Given a polynomial $f$ and a parameter $w$, the goal is to find an order $\sigma$ in which $f$ has an ROABP of \emph{width} $w$. We show that this problem is $\NP$-hard in the worst case, even when the input is a constant degree polynomial that is given in its dense representation.
We provide a reduction from $\cutwidth$ to prove these results.
Owing to the exactness of our reduction, all the known results for the hardness of approximation of $\cutwidth$ also transfer directly to the order-finding problem.
Additionally, we also show that any constant-approximation algorithm for the order-finding problem would imply a polynomial time approximation scheme (PTAS) for it.

On the algorithmic front, we design algorithms that solve the order-finding problem for generic ROABPs in polynomial time, when the width $w$ is polynomial in the individual degree $d$ of the polynomial $f$.
That is, our algorithm is efficient for most/random ROABPs, and requires more time only on a lower-dimensional subspace (or subvariety) of ROABPs.
Even when the individual degree is constant, our algorithm runs in time $n^{O(\log w)}$ for most/random ROABPs.
This stands in strong contrast to the case of (Boolean) ROBPs, where only heuristic order-finding algorithms are known.

\end{abstract}

\newpage

\pagenumbering{arabic}
\setcounter{page}{1}

\section{Introduction}
Read-Once Algebraic Branching Programs (ROABPs) are a well-studied and well-understood model in algebraic circuit complexity. There are numerous reasons for studying this model, starting with the fact that ROABPs serve as the algebraic analog of ordered ROBPs: Read-Once Branching Programs, a.k.a. OBDDs. Consequently, Polynomial Identity Testing (PIT) for ROABPs is the algebraic analog of the fundamental $\mathsf{RL}$ vs $\mathsf{L}$ problem in derandomization of boolean computation. 

Formally, an \textbf{ROABP} \( R(x_1,\ldots,x_n) \) for computing \( f(\vecx) \) in order \( \sigma \) is a layered, directed graph with \( n+1 \) layers. The source \( s \) and sink \( t \) are single vertices in the \(0^{\text{th}}\) and \(n^{\text{th}}\) layers, respectively. Edges between layers \( i-1 \) and \( i \) are labeled by univariate polynomials in \( x_{\sigma(i)} \) of degree at most \( d \). The polynomial computed by the ROABP is the sum of the products of edge weights along all paths from \( s \) to \( t \).

 The structural restriction of being read-once also makes ROABPs the commutative analog of non-commutative algebraic branching programs (ABPs). Despite this restriction, ROABPs can efficiently simulate many interesting algebraic models, such as sparse polynomials, set-multilinear depth-3 circuits, diagonal depth-3 circuits, and polynomials that have polynomially large dimension of partial derivatives (see e.g. \cite{FS13,BT24}). They can also efficiently compute complex polynomials like the Iterated Matrix Multiplication (IMM) polynomial~\cite{KNS15}, which is provably hard for constant-depth circuits~\cite{LST24}.

A major motivation for studying ROABPs is the characterization due to Nisan~\cite{N91} that precisely describes the ROABP size required for a given variable partition. This characterization has led to exponential lower bounds, a polynomial-time white-box PIT, and quasipolynomial-time black-box PITs for ROABPs and their variants (see e.g. \cite{GKST17}). Investigating the structure and properties of ROABPs has also led to significant advances in other areas of algebraic complexity. For instance, PITs for bounded top and bottom fan-in depth-4 circuits and ABP upper-bounds for the border of \(\Sigma \Pi \Sigma(k)\) circuits critically rely on the PIT and derandomization results for ROABPs (\cite{DDS21a,DDS21b}).

ROABPs have also proven to be highly useful for designing learning algorithms for various circuit classes, such as depth-3 powering (\( \sum \bigwedge \sum \)), set-multilinear depth-3 circuits, and polynomials with polynomially bounded partial derivatives~\cite{KS06}. Given an order and black-box access to a polynomial \( f \), we can (proper-)learn it in randomized polynomial time and deterministic quasipolynomial time. However, we currently do not know how to learn ROABPs without additional information about the ordering of variables. Therefore, it is extremely natural to ask whether we can design a learning algorithm that operates without this information of an optimal ordering. This question is the primary focus of our work. We begin by demonstrating the significant role that variable ordering plays in ROABPs.

\begin{fact}[See e.g. \cite{R06,FS13}]\label{fact:order-matters-roabp}
    The polynomial $f(x_1,\ldots,x_n,y_1,\ldots,y_n) := \prod_{i \in [n]} (x_i + y_i)$, has an ROABP of size $3n+1$ and width $2$ in the order $(x_1,y_1,x_2,y_2,\ldots,x_n,y_n)$, but requires ROABPs of size and width $2^{\Omega(n)}$ in the order $(x_1,\ldots,x_n,y_1,\ldots,y_n)$. 
\end{fact}

Given how the ROABP-size of a polynomial critically depends on the underlying ordering of the variables, let us formally state the problem of interest.

\begin{question}[Order finding for ROABP]\label{quest:ROABP-order-finding}
    Given a polynomial $f(x_1,\ldots,x_n)$ of individual degree $d$, and a number $w \in \N$, decide if there exists an order $\sigma \in s_n$ in which $f$ has an ROABP of width at most $w$. 
\end{question}
For concreteness, let us fix that the polynomial is given as either as a black-box, or as a circuit (``white-box'' representation).
Note that a \naive{} randomized algorithm for this problem would take \( n! \cdot \text{poly}(d,w) \) time by iterating over all the permutations, and checking whether for a particular partition \(\sigma\), the ROABP width: $\text{RO-width}_{\sigma}(f)$ is at most $w$.
But given the algebraic nature of the problem, and also the aforementioned characterization due to Nisan, it would be unfair to rule out a more tractable algorithm upfront.
This brings us to our results.

\subsection{Our Results}

Somewhat surprisingly, we show that the order-finding problem is indeed $\NP$-hard, by providing a polynomial-time Karp reduction from the $\cutwidth$ problem (see e.g. \cite{GJ79}) for linear arrangement of graphs.

\begin{theorem}
Given a polynomial \( f(x_1, \ldots, x_n) \) with individual degree \( d \), provided either as a black-box or a circuit, and a number \( w \in \mathbb{N} \), deciding whether there exists an order \( \sigma \in S_n \) such that \( f \) has an ROABP of width at most \( w \) is $\NP$-hard.
\end{theorem}

An important feature of this reduction is that it provides an \textit{exact} relationship between the cut-width of the input graph and the ROABP-width of the output polynomial. Concretely, we show that for any graph \( G \), we can construct a polynomial \( f_G \) such that \( \text{RO\text{-}width}_{\sigma}(f_G) = 2 + \cutwidth_{\sigma}(G) \).

Another important aspect of our construction is that for a graph of degree $\Delta$, \( f_G \) has a total degree bounded by \( 2 \Delta \). Combined with the known $\NP$-hardness results for cut-width for graphs with maximum degree 3, this shows that the hard instances for ROABP order-finding can even be constant-degree polynomials.

An interesting consequence of our hardness result for order finding is that the corresponding learning problem and the algebraic circuit minimization problem for ROABPs also become $\NP$-hard. Recall that in circuit minimization for any circuit class $\mathcal{C}$, we are given a polynomial (as an algebraic circuit) and a parameter \( s \), and asked if there exists a circuit \( C \in \mathcal{C} \) of size at most \( s \).

\begin{corollary}[Refer \autoref{thm:NP-hardness-worst-case-ckt}]\label{cor:NP-hardness-ckt-minimization-informal}
The algebraic circuit minimization problem for ROABPs is $\NP$-hard.
\end{corollary}

Given that the hardness of ROABP order finding is already encapsulated by instances of constant-degree polynomials, this directly implies the hardness of the algebraic minimum circuit size problem (MCSP) for ROABPs with constant-degree polynomials. Specifically, given a polynomial \( f \) of degree \( d = O(1) \) in the dense representation (analogous to the truth table representation in the Boolean setting), and a parameter $w$, deciding whether there exists an ROABP computing \( f \) of width $w$, is $\NP$-complete.   

\begin{corollary}[Refer \autoref{thm:NP-hardness-worst-case-dense}]
For a fixed \( d \), the algebraic MCSP  for ROABPs is $\NP$-hard.
\end{corollary}
For precise definitions  of these algebraic meta-complexity problems and corresponding results,  see subsection \ref{sec:algebraic-MCSP}. 

Due to the existence of hard instances in constant-degree polynomials, the above hardness result is similar to the hardness of set-multilinear depth-3 MCSP and depth-3 powering MCSP for degree-3 polynomials, which arises from the hardness of tensor decomposition (respectively) \cite{H90,MR13b,S16}. This contrasts with the recent hardness result of \cite{BDSS24} which crucially requires high-degree polynomials for their hardness.


\paragraph*{} 
Another outcome of the \textit{exact} relationship between cut-width and ROABP-width is that known hardness of approximation results for cut-width directly imply the hardness of \(\alpha\)-approximation for order finding in ROABPs. In this context, the objective is to test if there exists an ordering of variables such that the ROABP-width in that order is less than \(\alpha w\).

Assuming the Small Set Expansion (SSE) conjecture, it is known that \(\text{cutwidth}\) is hard to approximate within any constant factor greater than 1~\cite{APW12}. This directly implies that, under SSE, it is hard to approximate the order-finding problem.

We also provide evidence of the hardness of approximation for RO-order finding without relying on the strong assumption of SSE. Specifically, we show how to boost the hardness of approximation for RO-order from a small fixed constant to an arbitrary constant. Formally, we demonstrate that an algorithm capable of approximating order finding within any (fixed) constant can be converted into a PTAS for the order-finding problem.

\begin{theorem}[Refer \autoref{thm:constant-approx-to-PTAS-ckt}]\label{thm:approx-to-PTAS-informal}
    For any constant $\alpha$, an $\alpha$-approximation algorithm for finding an order that minimizes ROABP-width, implies a PTAS for the same problem.
\end{theorem}

Note that the above result, along with the non-existence of a PTAS for cut-width, would imply hardness of approximation (HOA) for RO-order unconditionally. Although we could not find such results for cut-width in the literature.
There are known results for the closely related problem of optimal linear arrangement~\cite{AMS11}, and we believe that similar results can also be adapted to cut-width.

On the algorithmic front, we design a randomized algorithm for solving the order-finding problem for ROABPs. The worst-case complexity of our algorithm is \(2^n \, \text{poly}(w, d)\), which is a significant improvement over the trivial \(n! \, \text{poly}(d,w)\) time required to check all possible orders. We also show that for generic (random) ROABPs, our algorithm runs in polynomial time and outputs a correct order with high probability. This means our algorithm is efficient for most ROABPs, requiring additional time only on a lower-dimensional subspace (sub-variety) of ROABPs. This stands in strong contrast to the case of Boolean ROBPs, where only heuristic order-finding algorithms are known~\cite{W00,MT98}.

\begin{theorem}[Informal version of \autoref{thm:generic-case-runtime} and \autoref{thm:random-case-runtime}]\label{thm:generic-random-runtime-informal}
    Over any large enough field $\F$, there is a randomized algorithm $R$ that when given a random (or generic) polynomial $f(\vecx)$ with an ROABP of width $w$ (in some unknown order) as a black-box, outputs an order which achieves that ROABP-width for $f$, with high probability.
    
    Further, the running time of this algorithm is $\poly(n,d,w)$ when $w = d^{O(1)}$, and quasi-polynomial ($n^{O(\log w)}$) when $d$ is constant.
\end{theorem}
 

\subsection{Ideas behind our proofs}

Owing to the simplicity and the linear algebraic nature of Nisan's characterization, the proofs of our results are fairly clean.
We have therefore tried to describe almost all the key ideas behind them here.

\subsubsection*{\textsf{NP}-hardness in the worst case}

As alluded to earlier, the motivation for looking at $\cutwidth$ as the candidate problem to show the $\NP$-hardness for order finding (rather, ROABP-width) comes from the hardness result for OBDD-size~\cite{BI96}, which provides a reduction from the optimal linear arrangement problem.

The particulars of our reduction are fairly natural, in that given a graph $G$ we wish to design a polynomial $f_G$ with the following property.
An optimal linear arrangement of $G$ should correspond to an optimal order for an ROABP of $f_G$, and a natural way to achieve that is to ensure that the size of each cut in the linear arrangement corresponds to the width in that layer for an optimal ROABP for $f_G$ in that order.

These criteria mean that the variables should correspond to the vertices of the graphs, and for each edge $(u,v)$ in the graph $G$, $f_G$ should include a ``gadget'' with the variables $x_u$ and $x_v$. Whenever a set $T \subseteq [n]$ includes exactly one of $u$ and $v$, each such edge-gadget should ideally increase the rank of the Nisan matrix $M^T(F)$ by a definite amount, so that the $\nisan{T}{\verline{T}}{f}$ then corresponds to the size of the cut induced by the vertices in $T$.

If we just use $x_u x_v$ as the gadget, it already gives us a unique monomial corresponding to that edge. But, in order to ensure that all edges incident on $x_u$ (that cross the cut) add to the rank by the same amount, we need to ensure that even the exponent of $x_u$ (and $x_v$) in the gadget retains some information about the edge it came from.
This can be achieved by the gadget being $x_u^v x_v^u$.
Finally, observing that $x_u$ only needs to have degree-of-$u$ many distinct exponents in the final polynomial $f_G$, takes us to the gadget $x_u^{n_u(v)} x_v^{n_v(u)}$, where $n_u(v)$ is a unique index for $v$ in the set of neighbours of $u$.
This lets us achieve an individual degree that is essentially the maximum degree of $G$, and a total degree that is just twice that.
This improvement is crucial in proving $\NP$-hardness of MCSP for ROABPs, in \autoref{cor:ROABP-MCSP-NP-hard}.

Beyond this point, the final reduction (\autoref{lem:cutwidth-to-order-finding}) follows after some minor tweaks that are necessary to handle some corner cases for the choice of the set $T$.

\subsubsection*{Order-finding in the average (or generic) case}

When designing an algorithm for order-finding, we assume that the width parameter $w$ is given as an input, and that the input polynomials are promised to have an ROABP in some order of width $w$.
This is without loss of generality, since we can just search for the correct $w$, using the proper-learning (or reconstruction) algorithms in the known-order setting (see \autoref{sec:generic-and-random-case} for details).

On a high level, our algorithm (\autoref{alg:order-finding}) tries to perform a greedy exploration of the space of all the possible sets $T$ that satisfy $\nisan{T}{\overline{T}}{f} \leq w$.
Since for a ``correct'' order $\sigma$, any set $P$ that is prefix of $\sigma$ does satisfy $\nisan{T}{\overline{T}}{f} \leq w$, the goal of this exploration is to find a sequence of `good' sets of sizes $1,2,\ldots,n-1$, which looks like a sequence of prefixes of some permutation $\sigma$.
Such a $\sigma$ should then be a valid order for a width-$w$ ROABP computing $f$.

Consider the space of all possible inputs to the order-finding algorithm as above.
Every input polynomial has an ROABP of width $w$ in some order, say that order is $\sigma$.
What would be bad for the greedy exploration above, is that a set $T$ that is not a prefix of a ``correct'' order for $f$ (which is possibly different from $\sigma$), still has $\nisan{T}{\overline{T}}{f} \leq w$.
Worse even, if several such sets exist, then the algorithm would have to explore the paths suggested by these sets, only to then abandon them when they fail.

Now the intuition behind why such an algorithm should run efficiently on a generic input is as follows.
Having small Nisan rank for a set $T$ that is not a prefix of $\sigma$ should be a ``special'' property, and so a \emph{generic} polynomial with a width-$w$ ROABP in the order $\sigma$, should be unlikely to have any such ``special'' properties.
Phrased in algebraic-geometric terms, this means that for any `bad' set $T$ and the \emph{variety} $V$ of polynomials that have a width-$w$ ROABP in the order $\sigma$, the polynomials that additionally have $\nisan{T}{\overline{T}}{f} \leq w$ form a sub-variety of $V$ of a strictly smaller dimension; this is precisely \autoref{lem:generic-ROABPs-bad-partitions}.
This then means that such polynomials form a tiny fraction of $V$, and thus, for a large enough (and especially an infinite) field, even the union of all the ``special'' sub-varieties corresponding to all the bad sets should still leave out most of the original variety; this is shown in \autoref{lem:bad-partition-varieties-dont-cover}.

All these arguments translate to analogous results over large finite fields, with careful applications of the Schwartz-Zippel lemma, to give us the running time guarantees.
A minor issue is that when the individual degree $d$ is too small compared to $w$, any set $T$ that is small enough trivially achieves a Nisan rank less than $w$.
This means that our exploration has to inevitably explore some `bad sets', which give us a super-polynomial running time when $d$ is constant.

\subsubsection*{PTAS from a constant approximation}

Our idea is to simply boost the approximation ratio.
This is achieved using a ``black-box'' transformation on polynomials (\autoref{lem:tensoring-roabp-width}), that maps any $n$-variate polynomial $f$ that has a width-$w$ ROABP in some order, to an $n$-variate polynomial $g_k$ that has a width-$w^k$ ROABP in the same order.
In fact, this transformation works for any $k$, and further, it raises the width in each layer to the $k^{th}$ exponent.
Here, by a ``black-box'' transformation, we mean that the transformation is oblivious to what $f$ is, what the order $\sigma$ is, or what the width $w$ is.

With this transformation, and a constant-approximation algorithm at hand, we obtain a PTAS as follows.
Say the approximation ratio of the algorithm is $\alpha$.
If we want an order that achieves a width that is at most $(1+\epsilon)$ times the optimal, we apply the transformation for $k \sim \log_{(1+\epsilon)} \alpha$, and run the $\alpha$-approximation algorithm on $g_k$.
We then output the exact same order given by the approximation algorithm, as the $(1+\epsilon)$-approximate order for $f$.
It turns out that this transformation is simple enough, so that the entire procedure runs in time $m^{O(\sfrac{1}{\epsilon})}$ on all inputs of length $m$, giving us a polynomial time approximation scheme.

\section{Related works}\label{sec:related-works}

\paragraph{Algorithms for ROABPs} 
Due to the strong structure of ROABPs, algorithms for various problems involving ROABPs have been extensively studied. For instance, we have a polynomial time PIT algorithm for ROABPs in the white-box setting \cite{RS05} and a  $(ndw)^{O(\log n)}$ (quasi-polynomial) time PIT algorithm in the black-box model \cite{FSS14, AGKS15}. Here, \( n \) is the number of variables, \( d \) is the individual degree, and \( w \) is the width. Furthermore, there are faster $(ndw)^{O(\log \log w)}$ black-box PIT algorithms for subclasses of ROABPs that are order-oblivious (a.k.a. commutative) ROABPs \cite{GKS17}.

Additionally, when provided with black-box access to a polynomial and a variable ordering, learning ROABPs becomes a tractable problem. Formally, given a variable ordering, an ROABP can be learned in randomized polynomial time and deterministic quasi-polynomial time, specifically \( (ndw)^{O(\log n)} \)~\cite{KS06, FS13}.

\paragraph{Ordered ROBPs} The ROABP class is inspired by the Boolean circuit class of ordered read-once branching programs (ROBPs), also referred to as Ordered Binary Decision Diagrams (OBDDs). These classes have been extensively studied due to their connection to the $\mathsf{RL}$ vs $\mathsf{L}$ problem \cite{N90} which is the central question in space-bounded derandomization.

An ordered ROBP, or OBDD, is a branching program in which the variables are read exactly once in a fixed, consistent order across all paths from the source to the sink. The program represents Boolean functions as directed acyclic graphs, where nodes correspond to variables, and edges represent the two possible outcomes for each variable.

Even for ordered ROBPs, the problem of \textit{order finding}—that is, given an ordered ROBP, determining whether there exists an ordering with ROBP width at most \( w \)—is a natural computational problem. In fact, this problem has been extensively studied in the Boolean setting.

Numerous works address both the hardness and algorithms for exact as well as approximate \textit{order finding} for ROBPs. For instance, it is known that the order finding for OBDDs is $\NP$-hard \cite{BI96} and even $\NP$-hard to approximate for any constant factor  \cite{S02}.  Despite these hardness results, and due to the importance of OBDDs in both theoretical and practical fields—such as VLSI design, formal verification, machine learning, and combinatorial problems—numerous \emph{heuristic} algorithms have been developed to solve the order-finding problem for OBDDs, see \cite{W00,MT98}. 
For a comprehensive survey about the study of ordering in OBDDs, we refer the reader to the beautiful and well detailed book  by Wegner \cite{W00}.

Although ROABPs share an analogous criterion for width as ROBPs, none of these Boolean results directly imply hardness or algorithmic insights for order-finding in the ROABP setting. This is because the standard algebraization of a Boolean function can yield significantly different complexities in ROABPs compared to ROBPs.
The key reason for this difference is that the result analogous to Nisan's characterization (\autoref{thm:Nisan-characterization}) for OBDDs relies on the \emph{number} of distinct sub-functions~\cite{B86}, whereas Nisan's result is based on their rank (or the dimension of the space spanned by them). We illustrate this difference with a small example below.

\begin{example}\label{eg:obdd-vs-roabp}
    Consider the function $f(a,b,c,d)$ that evaluates to $1$ exactly on the following inputs.
        $$I = \set{(0,0,0,0),(0,1,0,1),(1,0,1,0),(1,1,0,0),(1,1,1,0)}$$
    
    Any OBDD for $f$ in the order $(a,b,c,d)$ requires width $4$ in the second layer, whereas (when viewed as a multilinear polynomial) $f$ has an ROABP in the order $(a,b,c,d)$ that has width $3$ in the second layer.
\end{example}

Nonetheless, these results serve as valuable guides for what we might expect in the algebraic setting as well. In fact, our use of graph linear arrangement and cutwidth was inspired by reductions for the hardness of order-finding in ROBPs from \cite{BI96}. The idea of boosting a small, constant hardness to an arbitrary constant hardness also originates in work by \cite{S02}. However, in both of these reductions, we can structure the algebraic instances such that the reductions are more efficient and conceptually simpler.

It is also worth noting that no rigorous algorithm is  known for finding the order for generic or random OBDDs, indicating some key differences here.

\paragraph{Other $\NP$-Hardness Results in Algebraic Complexity}
Unlike the Boolean world, only limited $\NP$-hardness results are known for problems involving algebraic circuits and algebraic computations. Firstly, there are hardness results regarding (proper) learning of set-multilinear depth-3 circuits and depth-3 powering circuits. These results follow directly from the hardness of computing tensor rank (and Waring rank) \cite{H90, S16}, along with standard connections between tensor rank and set-multilinear depth-3 circuits, and Waring rank with depth-3 powering circuits \cite{BSV21}.

Additionally, there are known hardness results for equivalence testing, which essentially asks if two polynomials are equivalent up to a change of basis. Kayal \cite{K11}, and Agrawal-Saxena \cite{AS06}   showed that this problem is as hard as the graph isomorphism problem even for cubic polynomials!

Recently, \cite{CGS23} demonstrated the $\NP$-hardness of testing whether there exists a shift of a polynomial that is sparser than the polynomial itself. They showed that this problem is $\NP$-hard over finite fields and undecidable over the integers.

Even more recently, \cite{BDSS24} showed that, given a sparse polynomial, testing whether it is equivalent to an \(s\)-sparse polynomial under an invertible change of basis is $\NP$-hard. They also showed that this problem is $\NP$-hard to approximate up to $s^{\frac{1}{3}}$-approximation factor.

\paragraph{Generic Learning Algorithms in Algebraic Complexity} 
It is widely believed that learning (or reconstruction) problems for most reasonably strong classes of algebraic circuits are hard. Consequently, there has been considerable effort in designing learning algorithms for the generic case. This focus is partly due to connections with important problems, such as tensor decomposition, subspace clustering, and learning mixtures of Gaussian \cite{GKS20, CGKMS24}. In the generic case, we assume that the input comes from the space excluding some strictly smaller-dimensional variety. Naturally, generic algorithms also give average-case algorithms, as a randomly sampled point will avoid the subvariety with high probability.

We know of generic learning algorithms for several constant depth circuit models such as homogeneous depth-3, set-multilinear depth-3, generalized depth-3  circuits \cite{KS19b, BGKS22} and also 'sum of power of low-degree polynomials' \cite{GKS20}. The main idea, as established by the work of Kayal and Saha \cite{KS19b}, is to utilize the space of partial derivatives (or shifted partial derivatives), which has been used to prove circuit lower bounds, and to apply vector decomposition algorithms on these spaces.

Notably, our learning algorithm for generic ROABPs diverges from this framework, since any generic algorithm for learning ROABPs would primarily require a general approach for order finding. Our work examines the relationship between various permissible orderings in generic ROABPs.

\section{Preliminaries}

\subsubsection*{Notation}
Apart from the conventions of notation that are commonly used, we follow these additional conventions in our paper.  
\begin{itemize}
    \item We use lowercase letters to denote indeterminates, polynomials or scalars from the field, and use the boldfaced, lowercase versions of the same letters to denote sets or vectors of the corresponding things. E.g. $\vecx = \set{x_1,\ldots,x_n}, \veca = \set{a_1,\ldots,a_n}$.

    We use subscripts on these symbols for variable or polynomial sets to denote a subset. E.g. $\vecx_T = \set{x_i : i \in T}$.
    
    \item We use uppercase letters to denote matrices (and numbers that are very large), and use $A[i,j]$ to denote the $(i,j)$th entry of the matrix $A$.
    \item We use calligraphic uppercase letters for random variables, and use the regular version of the same letters, lower or uppercase (but never both), to denote their values or realizations. E.g. the random variables $\mathcal{A}$ and $\mathcal{P}$ take values $A$ and $p$.

    \item For a monomial $m$, the \emph{support of $m$} refers to the number of distinct variables that appear in the monomial: e.g. $xy$ has support $2$, $x^{100}$ has support $1$.

    \item For a polynomial, its total degree is the maximum degree of any monomial appearing in it, and its individual degree is the highest power that any individual variable is raised to in any monomial appearing in it.
    For instance, an $n$-variate multilinear polynomial has individual degree $1$, and total degree (possibly) $n$.

    \item For a polynomial $f$, the \emph{coefficient vector of $f$} is a list of the coefficients of all its monomials (even those that are zero) in some predefined order. For instance, for an $n$-variate, total-degree $d$ polynomial, the coefficient vector has length $\binom{n+d}{d}$, and an $n$-variate, individual-degree $d$ polynomial has a coefficient vector of length $(d+1)^n$.
\end{itemize}

\paragraph*{Linear arrangement problems on graphs.}
Given a graph $G$ on $n$ vertices, a linear arrangement of $G$ is a mapping from the vertex set to the points $\set{1,\ldots,n}$, or, just a permutation of the vertices.
For a linear arrangement $\sigma$ of a graph $G$, one can then look at various graph parameters.
One such parameter is the number of edges crossing each of the ``induced cuts'': $\set{\sigma^{-1}(1),\ldots,\sigma^{-1}(i)}$ and $\set{\sigma^{-1}(i+1),\ldots,\sigma^{-1}(n)}$ for each $i$.
The \emph{optimal} or \emph{minimum linear arrangement} (OLA or MLA) asks for an arrangement that minimizes the sum of the sizes of all these cuts, and the $\cutwidth$ problem asks for an arrangement that minimizes the sizes of the largest such cut.
We refer the reader to the survey by {D\'{\i}az}, Petit and Serna~\cite{DPS02} for an overview of some well-studied linear arrangement problems on graphs.

\subsection{Basics about polynomials and ROABPs}

Our proofs repeatedly use the following famous fact about multivariate polynomials, which has been proven in various guises in several works over the years~\cite{O22,DL78,S80,Z79}.
\begin{lemmawp}[Schwartz-Zippel Lemma]\label{lem:schwartz-zippel}
    Let $\F$ be any field, let $S \subseteq \F$ be arbitrary, and suppose $f(\vecx)$ is an $n$-variate polynomial with total degree $d$.
    Then, for independent $a_1,a_2,\ldots,a_n$ sampled uniformly at random from $S$, $\Pr[f(a_1,\ldots,a_n) = 0] \leq \sfrac{d}{\abs{S}}$.
\end{lemmawp}

\begin{definition}[Read-once Oblivious ABP (ROABP)]\label{defn:ROABP}
    Let $\F$ be any field.
    Let $n,d \in \N$ be arbitrary, let $\sigma \in s_n$ be a permutation, and consider an $n$-variate polynomial $f(\vecx)$ of individual degree $d$.

    An ROABP $R(\vecx)$ computing $f(\vecx)$ in the order $\sigma$ is a layered, directed graph with $n+1$ layers of vertices, with the leftmost ($0^{th}$) and the rightmost ($n^{th}$) layers having single vertices, called the source $s$ and the sink $t$.
    All edges in the graph are only between layers $i-1$ and $i$ for $i \in [n]$ labelled from left to right.
    Further, the weights on all the edges between layers $i-1$ and $i$ are univariate polynomials of degree at most $d$ in the variable $x_{\sigma(i)}$.

    Similar to an algebraic branching program (ABP), the polynomial computed by the ROABP is the sum of the polynomials computed along the different $s$ to $t$ paths, where each path computes the products of the weights of the edges contained in it.

    The \emph{size} of an ROABP is said to be the total number of vertices in it, and the \emph{width} of an ROABP is the maximum number of vertices it has in any layer. 
\end{definition}

Using the adjacency matrices of the sets of edges between all pairs of consecutive layers of an ROABP, we get an equivalent matrix view defined below.

\begin{definition}[ROABP (matrix view)]\label{defn:ROABP-matrices}
    Let $\F$ be any field.
    For any $n,d,w \in \N$, a permutation $\sigma \in s_n$, and an $n$-variate polynomial $f(\vecx)$ of individual degree $d$, we say that it has a width $w$ ROABP in the order $\sigma$, if there exist matrices $\set{A_{i,j}}$ with entries in $\F$ for all $i \in [n]$ and $0 \leq j \leq d$, such that the following holds.
    
    For $w_0 = w_n = 1$, there are numbers $w_1,w_2,\ldots,w_{n-1} \leq w$, so that for each $i \in [j]$, $A_{i,j} \in \F^{w_{i-1} \times w_i}$ for all $j \in \set{0,1,\ldots,d}$, and
    \begin{align*}
        f(\vecx) &= \inparen{L_{\sigma(1)}(x_{\sigma(1)}) \cdot L_{\sigma(2)}(x_{\sigma(2)}) \cdots L_{\sigma(n)}(x_{\sigma(n)})}[1,1],\\
        &\text{where for all $i \in [n]$,}\\
        L_{i}(x_{i}) &= A_{i,0} + A_{i,1} x_i + A_{i,2} x_i^2 + \cdots + A_{i,d} x_i^d.
    \end{align*}
    We call the matrices $\set{A_{i,j}}$ the \emph{coefficient matrices} of the ROABP.
\end{definition}

\begin{definition}[Prefixes of a permutation]\label{defn:prefix-of-permutation}
    For a permutation $\sigma \in s_n$ on $n$ letters, we say that a set \emph{$T \subseteq [n]$ is a prefix of (or with respect to) $\sigma$} if $T = \set{\sigma(1),\sigma(2),\ldots,\sigma(i)}$ for some $i \in [n]$.
\end{definition}

\begin{definition}[Nisan matrix]\label{defn:nisan-matrix}
    For an $n$-variate polynomial $f(\vecx) \in \F[\vecx]$ of individual degree $d$, and a subset $T \subseteq [n]$, the \emph{Nisan matrix of $f$ with respect to $T$} is a $(d+1)^{\abs{T}} \times (d+1)^{n - \abs{T}}$ matrix denoted by $M_T(f)$, with rows labelled by monomials in the variables $\vecx_{T}$, and columns labelled by the monomials in $\vecx_{[n] \setminus T}$, and the $(m,m')$th entry of this matrix is $M_T(f)[m,m'] = \coeff_{f}(m \cdot m')$.
\end{definition}

\begin{theoremwp}[Nisan's characterization~\cite{N91}]\label{thm:Nisan-characterization}
    Let $\F$ be any field.
    For any $n \in \N$ and any permutation $\sigma \in s_n$, let $\emptyset,T_1,T_2,\ldots,T_{n-1},[n]$ be the prefixes of $\sigma$ of lengths $0,1,2,\ldots,n-1,n$, respectively. 
    
    Then for any $n$-variate polynomial $f(\vecx) \in \F[\vecx]$, the optimal ROABP for $f$ in the order $\sigma$ has 
    exactly $\nisan{T_i}{\bar{T_i}}{f}$ vertices in the $i$th layer (after ``reading'' first $i$ variables).
    Thus, the optimal ROABP has width exactly $\max_{i \in [n]} \nisan{T_i}{\bar{T_i}}{f}$, and size exactly $\sum_{i \in [n]} \nisan{T_i}{\bar{T_i}}{f}$.
\end{theoremwp}

\subsection{Generic and random ROABPs}\label{subsec:generic-and-random-ROABPs}

\textbf{Generic Conditions}: A generic point of an \emph{algebraic variety} $V \subseteq \mathbb{F}^N$ is one that lies outside any proper closed subset of $V$ defined by additional algebraic conditions. In other words, it avoids the "special" sub-varieties where additional polynomial relations hold, meaning it satisfies the "general" conditions that are typical of most points on $V$.

Thus, for $\F$ being either the field of rational, real or complex numbers, we consider the variety formed by the \emph{coefficient vectors} of length $(d+1)^n$ of all $n$-variate, individual degree $d$ polynomials that have an ROABP of width $w$ in some order $\sigma$ (see \autoref{subsec:complexity-generic-case}).
Thus, by a generic ROABP (of a certain width in a certain order), we simply mean that the coefficient vector is generic and does not come from a strict subvariety of the above variety. It turns out that over any infinite field and for some well-behaved varieties, if we sample a random ROABP, it will be a generic ROABP with high probability.
We formalize the notion of a random ROABP below (as it is more amenable for computer science applications). 

\begin{definition}[Random ROABP]\label{defn:random-ROABP}
    Let $\F$ be any field, and let $n,d,w \in \N$, $\sigma \in s_n$, and $S \subseteq \F$ be arbitrary.
    Define random variables $\vec{\mathcal{A}} = \set{\mathcal{A}_{i,j}}_{i \in [n], j \in \set{0,1,\ldots,d}}$, where each $\mathcal{A}_{i,j} \in \F^{w \times w}$ each of whose entries is sampled independently and uniformly at random from $S$.
    This then gives us a random variable $\mathcal{P}_{S,n,d,w,\sigma}$, which is the polynomial computed by the ROABP given by
    \[
        \mathcal{P}_{S,n,d,w,\sigma}(x_1,\ldots,x_n) := \inparen{\prod_{i \in [n]} \inparen{\sum_{j = 0}^{d} \mathcal{A}_{i,j} \cdot x_{\sigma(i)}^j}}[1,1].
    \]
    We call this random variable $\mathcal{P}_{S,n,d,w,\sigma}$ a \emph{random $n$-variate ROABP of individual degree $d$ of width $w$ in the order $\sigma$ over the set $S$}.
    That is, the distribution on the realizations of this random variable is induced by the uniform distribution over $S$ in the entries of all the $\mathcal{A}_{i,j}$s.

    We drop the subscripts $S,n,d,w,\sigma$ and just write $\mathcal{P}$ when those details are clear from the context.
\end{definition}

\begin{remark}\label{rmk:sampling-random-ROABPs}
    Note that \autoref{defn:random-ROABP} can be defined analogously for other ``standard'' distributions the entries of the coefficient matrices. The only criterion is that these distributions should follow an analogue of the Schwartz-Zippel lemma \autoref{lem:schwartz-zippel}.
\end{remark}

\begin{observation}\label{obs:degree-of-coeffs-of-P-in-As}
    For any valid $\F,\sigma,n,d,w$, the coefficients of $\mathcal{P}_{n,d,w,\sigma}$ are homogeneous multilinear polynomials in the random variables $\vec{\mathcal{A}}$ of total degree exactly $n$.
\end{observation}
\begin{proof}
    The claim follows by observing that for any monomial $m = x_1^{e_1} x_2^{e_2} \ldots x_n^{e_n}$, its coefficient in $\mathcal{P}$ is exactly
    \[
        \coeff_{\mathcal{P}}(m) = \inparen{\prod_{i \in [n]} \mathcal{A}_{i,e_i} } [1,1]. \qedhere
    \]
\end{proof}

\subsection{Useful facts about ROABPs}\label{subsec:roabp-facts}

We start by stating the following facts about Nisan matrices, which are essentially facts about the ranks of Kronecker products and sums of matrices.
\begin{observationwp}[Properties of Nisan matrices]\label{obs:properties-nisan-matrices}
    For polynomials $f(x_1,\ldots,x_n), g(y_1,\ldots,y_n)$ and subsets $T,U \subseteq [n]$, we have the following.
    \begin{enumerate}
        \item For $h(z_1,\ldots,z_{2n}) := f(z_1,\ldots,z_n) \cdot g(z_{n+1},\ldots,z_{2n})$ and $U' := \set{n+u : u \in U}$,\\
              $\nisan{T \cup U'}{\overline{T \cup U'}}{h} = \nisan{T}{\overline{T}}{f} \cdot \nisan{U}{\overline{U}}{g}$.
        \item For $h(z_1,\ldots,z_n) := f(z_1,\ldots,z_n) + g(z_{1},\ldots,z_n)$,\\
              $\nisan{T}{\overline{T}}{h} \leq \nisan{T}{\overline{T}}{f} + \nisan{T}{\overline{T}}{g}$. \qedhere
    \end{enumerate}
\end{observationwp}

\begin{lemma}[Evaluation dimension and Nisan matrix]\label{lem:eval-dim-nisan-rank}
    Suppose $f(\vecx) \in \F[\vecx]$ be an $n$-variate polynomial with individual degree $d$.
    Let $T \subseteq [n]$ be any subset and $S \subset \F$ with at least $(d+1)$ elements.
    Define a matrix $E_T(f)$ with $(d+1)^{\abs{T}}$ rows and $(d+1)^{(n - \abs{T})}$ columns, indexed by \emph{evaluation points} from $S^{\abs{T}}$ and $S^{(n - \abs{T})}$, respectively; the entries of $E_T(f)$ are: $E_T(f)[\alpha,\beta] = f(\vecx_{T} = \alpha, \vecx_{([n] \setminus T)} = \beta)$.
    
    Then, $\mathrm{rank}(E_T(f)) = \nisan{T}{[n] \setminus T}{f}$.
\end{lemma}
\begin{proof}[Proof sketch]
    Let $S = \set{s_1,\ldots,s_{(d+1)}}$, and suppose $V_{S}$ is the $(d+1) \times (d+1)$ Van der Monde matrix, with $V[i,j] = s_i^{j-1}$.
    Then the proof follows from the fact that $E_T = V_S^{\otimes \abs{T}} \cdot M^{T} \cdot \inparen{V_S^{\mathsf{T}}}^{\otimes \abs{T}}$, where $A^{\otimes k}$ is the $k$-wise Kronecker/tensor product of the matrix $A$ with itself.
\end{proof}

\begin{lemma}[Estimating Nisan rank using PIT]\label{lem:nisan-rank-estimation-from-PIT}
    Let $n,d,w \in \N$ be arbitrary, and let $S \subseteq \F$ be any set of field constants.
    Suppose $f(\vecx)\in\F[\vecx]$ is an $n$-variate polynomial of individual degree $d$, and $T \subseteq [n]$ is of size $k$.

    Then for $\alpha_1,\alpha_2,\ldots,\alpha_{w+1} \in S^{k}$ and $\beta_1,\beta_2,\ldots,\beta_{w+1} \in S^{n-k}$ drawn uniformly at random, the $(w+1) \times (w+1)$ matrix $\hat{E}$ with $\hat{E}[i,j] = f(\vecx_{T} = \alpha_i, \vecx_{[n] \setminus T} = \beta_j)$ for each $i,j \in [w+1]$, has a nonzero determinant if and only $\nisan{T}{\overline{T}}{f} > w$, with probability at least $\sfrac{nd(w+1)^2}{\abs{S}}$. 
\end{lemma}
\begin{proof}
    Let us define a $(w+1) \times (w+1)$ matrix $E_0$, with polynomials in the ``fresh'' variables $\vecy_1,\ldots,\vecy_{w+1}$ and $\vecz_1,\ldots,\vecz_{w+1}$ as entries.
    In particular, $E_0[i,j] = f(\vecx_{T} = \vecy_i, \vecx_{[n] \setminus T} = \vecz_j)$ for each $i,j \in [w+1]$.
    The determinant of this matrix is a nonzero polynomial, \emph{if and only if} $\nisan{T}{\overline{T}}{f} > w$, because we can substitute the $\vecy_i$s and the $\vecz_j$s appropriately, from \autoref{lem:eval-dim-nisan-rank}.
    Further, the total degree of this determinant is $\deg(f) \cdot (w+1)^2 \leq nd \cdot (w+1)^2$.

    Therefore, checking whether the $\nisan{T}{\overline{T}}{f} > w$ reduces to the PIT of this determinant.
    By the Schwartz-Zippel lemma (\autoref{lem:schwartz-zippel}), this is the same as testing if the determinant of the random matrix $E$ is nonzero, as claimed.
\end{proof}

\section{Worst case \textsf{NP}-hardness}\label{sec:worst-case-complexity}

This section is devoted to proving the $\NP$-hardness of \autoref{quest:ROABP-order-finding} in the worst case as stated below.
When the total degree is constant, we obtain an $\NP$-hardness result for testing ROABP width, even when the polynomial is provided in its \emph{dense representation}\footnote{Dense representation of $f$ lists only the coefficients of all the monomials of $f$ in some pre-determined order; such a representation is suited for a polynomial that is \emph{dense}: has many monomials. On the other hand, \emph{sparse representation} of $f$ lists pairs of monomials and their coefficients, which is ideal for a sparse polynomial.}.
We also get an analogous result for when the degree grows with $n$, the only difference being that then the dense representation has size that is $n^{\omega(1)}$, and hence we have to switch to the \emph{sparse representation}.
We now formally define the corresponding languages.

\begin{definition}[Problems around finding ROABP-width]\label{defn:width-finding-problems}
    For every positive integer $d$, we define the language $\DenseROwidth{d}$, to contain tuples $(\cvector(f),w)$, where $\cvector(f)$ is the vector of the $\binom{n+d}{d} = n^{O(d)}$ coefficients of an $n$-variate, total-degree-$d$ polynomial $f$, and $w \in \N$, such that $f$ has an ROABP of width $w$ in some order.

    We define the language $\CktROwidth$ to contain $4$-tuples $(C,n,d,w)$, where $n,d \in \N$ are given in \emph{unary}, $w \in \N$ is given in binary, and $C$ is the description of an $n$-variate, degree-$d$ polynomial $f$ represented as an algebraic circuit, such that $f$ has an ROABP of width $w$ in some order.

    The search version of these problems: $\mathtt{Search}$-$\DenseROwidth{d}$ and $\mathtt{Search}$-$\CktROwidth$ are defined analogously, where the parameter $w$ is not provided as an input, and an algorithm solving the problem is expected to output an order $\sigma$ that achieves the optimal ROABP-width. 
\end{definition}

\begin{theorem}\label{thm:NP-hardness-worst-case-dense}
    For any integer $d \geq 6$, the language $\DenseROwidth{d}$ is $\NP$-hard under polynomial-time Karp reductions.
\end{theorem}

In order to prove the above theorem, we provide a polynomial-time (Karp) reduction from the $\cutwidth$ problem, defined for graphs (see e.g. \cite{DPS02}).
An important feature of this reduction is that it provides an exact relationship between the cut-width of the input graph and the ROABP-width of the output polynomial.

\begin{lemma}[Reduction from \texttt{CutWidth}]\label{lem:cutwidth-to-order-finding}
    Given a graph $G = (V,E)$ with $n$ vertices, $m$ edges, and maximum degree $\Delta$ there is a deterministic polynomial-time algorithm that outputs a polynomial $f(x_1,\ldots,x_n)$ of individual degree $d = \Delta + 1 $ with $n+m$ monomials, with the following property.
    
    There is an ROABP of width $w+2$ in an order $\sigma \in s_n$ \emph{if and only if} the linear arrangement of $G$ in the order $\sigma$ has $\cutwidth$ $w$. 
\end{lemma}
\begin{proof}
    Fix $V=[n]$. For every  $u \in V$, let  $n_u(v) \in [\Delta]$ denote a number that uniquely identifies $v$  within the neighborhood of $u$ (denoted $\text{Nbr}(u)$).
    One canonical way of fixing $n_u(v)$ is just by fixing an order on the vertices, and then for every $u \in V$ defining $n_u(v)$ in increasing order.

    For $i \in [m]$ and $e_i = \set{u,v} \in E$, define the monomial $f_i(x_u,x_v) = x_u^{n_u(v)} \cdot x_v^{n_v(u)}$, and let
    $$f_G(\vecx) := \sum_{v \in [n]} x_v^{\Delta +1} + \sum_{i \in [m]} f_i  = \sum_{i \in [n]} x_i^{\Delta +1} + \sum_{i \in [m]} x_u^{n_u(v)}x_v^{n_v(u)}.$$
    Note that $f_G$ has a monomial for each edge along with an $\sum_{i \in [n]} x_i^{\Delta +1}$ `gadget`. Its support size is constant ($2$),  individual degree is $\Delta+1$, and total degree is $2 \Delta $.

    The proof of the lemma follows directly from the following claim using Nisan's characterization (\autoref{thm:Nisan-characterization}). 
    \begin{claim}
        For any non-trivial partition $A \sqcup B = [n]$, we have $\nisan{A}{B}{f_G} = 2 + c_G(A,B)$ where $c_G(A,B)$ is the number of edges in $G$ that have one endpoint in $A$ and another in $B$.
    \end{claim}
    \begin{proof}
        \begin{description}
            \item[\underline{$\rank(M^{A,B}_{f_G}) \leq 2 + c_G(A,B)$}:] The bound follows from the following description of $f_G$ as a sum of $2+c_G(A,B)$ variable disjoint (w.r.t $A, B$) products.
            \begin{align*}
                f_G(\vecx) &= \sum_{e_i = \set{u,v} \subset A}  x_u^{n_u(v)} x_v^{n_v(u)} + \sum_{v \in A} x_v^{\Delta + 1}  \\
                         &+ \sum_{e_j = \set{u,v} \subset B} x_u^{n_u(v)} x_v^{n_v(u)} + \sum_{v \in B} x_v^{\Delta + 1} \\
                         &+ \sum_{\substack{e_k = \set{u,v}\\u \in A, v \in B}} x_u^{n_u(v)} x_v^{n_v(u)}.
            \end{align*}
            That is, the first and second terms contribute rank 1 each, and the last term contributes at most $c_G(A,B)$.

            \item[\underline{$\rank(M^{A,B}_f) \geq 2 + c_G(A,B)$}:] To show this, we exhibit a permutation matrix of dimension $c_G(A,B)$ that is a submatrix of $M^{A,B}(f)$, along with $2$ rows that lie in a disjoint space.
            First, let us partition the edges into three sets; $L$ : edges with both endpoints in $A$, $R$ : edges with both endpoints in $B$, and $C$ : edges that have exactly one endpoint in each $A$ and $B$.
            
            Now consider the submatrix defined by the following: for each edge $e_i = \{u,v\} \in C$, select the \emph{row} indexed by $x_u^{n_u(v)}$ and the \emph{column} indexed by $x_v^{n_v(u)}$. Note that, by construction, for each row in this submatrix, there is \emph{exactly} one column with a non-zero entry. The corresponding coefficient is also 1, thus forming a permutation matrix.

            Further, the row corresponding to any other monomials in $f_G$: coming from edges in $L$, or $x_i^{\Delta + 1}$ for $i \in A$, have the entry $1$ (in $M^{A, B}(f_G)$) corresponding to the constant polynomial in the column index, and zeroes in all other columns.
            Similarly, the columns indexed by edges in $R$ or $x_i^{\Delta + 1}$ for $i \in B$, have a $1$ (in $M^{A, B}(f_G)$) in row indexed by the constant polynomial, and zeroes everywhere else.
            Thus, those monomials cumulatively give an additional $2$-dimensional row-space disjoint from the $c_G(A,B)$-dimensional row-space that we get from the permutation submatrix discussed above. \qedhere
        \end{description}
    \end{proof}

    Clearly, $f_G$ has total degree $2 \Delta(G)$, has exactly $m+n$ monomials, and since all the nonzero coefficients are $1$, its dense representation has size $n^{O(\Delta(G))}$, and its sparse representation has size at most $\poly(n,m)$. Further, both these representations can easily be computed in time $\poly(n,m)$ from the graph $G$.
\end{proof}

\begin{proof}[Proof of \autoref{thm:NP-hardness-worst-case-dense}]
    With the reduction outlined in \autoref{lem:cutwidth-to-order-finding}, the theorem follows from the fact that $\cutwidth$ is $\NP$-hard under polynomial-time Karp reductions for planar graphs with maximum-degree $3$~\cite{MS88b}, as the dense representation then has size $\poly(n)$.
\end{proof}

Owing to the exactness of our reduction from the Cut-width problem, even when the total degree of the polynomial is growing (precisely when the graph $G$ has degree $\omega(1)$), the polynomial $f_G$ has the appropriate ROABP-width.
However, it is no longer possible to output the dense representation in time $poly(n)$ just because its length would be $n^{\omega(1)}$.
Thus, we only get the $\NP$-hardness result for the sparse representation, as stated below.

\begin{theorem}\label{thm:NP-hardness-worst-case-ckt}
    The language $\CktROwidth$ is $\NP$-hard under polynomial-time Karp reductions.
\end{theorem}
\begin{proof}
    As the polynomial $f_G$ generated by the reduction in \autoref{lem:cutwidth-to-order-finding} has at most $n+m$, support-$2$ monomials of individual degree at most $d = \Delta(G)+1 \leq n+1$ with all nonzero coefficients being just $1$, it can be represented as a sum of monomials with $O((n+m) \cdot 2 \log d) = O(n^2 \log n)$ bits.
    The theorem is then a direct result of $\cutwidth$ being $\NP$-hard for general undirected, unweighted graphs.
\end{proof}

\subsection{ Algebraic MCSP}\label{sec:algebraic-MCSP}

The Minimum Circuit Size Problem (MCSP) is a decision problem in which the input consists of the truth table of an \( n \)-variate Boolean function and a parameter \( s \in \mathbb{N} \). The goal is to decide whether \( f \) is computable by a circuit of size at most \( s \). Whether MCSP for general Boolean circuits is \(\mathsf{NP}\)-hard remains a long-standing open question with intriguing connections to cryptography, learning theory, average-case complexity, and proof complexity. For more details, see, for instance, \cite{H22, I21} and the references therein.


For restricted circuit classes \(\mathcal{C}\), one can define \(\mathcal{C}\)-MCSP, which asks whether, given a Boolean function (as a truth table) and \( s \in \mathbb{N} \), there exists a circuit \( C \in \mathcal{C} \) of size at most \( s \). It is known that MCSP is \(\mathsf{NP}\)-hard (under deterministic polynomial-time reductions) for DNF \cite{M79} and DNF \( \circ \) XOR formulas \cite{HOS18}. However, no \(\mathsf{NP}\)-hardness result is known (under deterministic polynomial-time reductions) for more general circuit models such as \(\mathsf{AC}^0\) circuits.

We now describe the minimum circuit size problem (MCSP) in the algebraic setting. Instead of a truth table for a Boolean function, you are given the complete coefficient vector of a polynomial. Specifically, for an \( n \)-variate polynomial of total degree \( d \), you are provided with all \( \binom{n+d}{d} \) coefficients. Alternatively, one could define the input as the evaluations of the polynomial at an interpolating set; note that it is possible to convert between these representations in polynomial time.

Furthermore, as in the Boolean setting, we can specialize this question to specific circuit classes \( \mathcal{C} \). Here, the problem is to determine if there exists a circuit of size \( s \) in \( \mathcal{C} \).

Below, we provide a formal definition for any algebraic circuit class \( \mathcal{C} \).

\begin{definition}[Algebraic MCSP for \( \mathcal{C} \)]\label{defn:algebraic-MCSP}
   Given an \( n \)-variate polynomial \( f \) of degree \( d \) along with its coefficient vector of \( \binom{n+d}{d} \) field coefficients and a parameter \( s \in \mathbb{N} \), decide if there exists an algebraic circuit \( C \) in \( \mathcal{C} \) of size \( s \) that computes \( f \).
\end{definition}

Another related problem in meta-complexity is the Circuit Minimization Problem. The key difference between circuit minimization and MCSP is that in circuit minimization, you are provided with an explicit representation of the function (a circuit computing \( f \)) that is usually polynomial in \( n \) size, and you are asked if there exists a circuit computing \( f \) with size less than \( s \). See, for instance, \cite{BU11} and the references therein. 

We can similarly define the algebraic version of the circuit minimization problem.

\begin{definition}[Algebraic Circuit Minimization for \( \mathcal{C} \)]\label{defn:algebraic-ckt-minimization}
   Given an \( n \)-variate polynomial \( f \) of degree \( d \), a circuit of size \( s' \) computing \( f \), and a parameter \( s \in \mathbb{N} \), decide if there exists an algebraic circuit \( C \) in \( \mathcal{C} \) of size \( s \) that computes \( f \).
\end{definition}

Note that the input size here is \( \text{poly}(n, d, s') \). It is intuitive to think that the circuit minimization problem may be harder than MCSP, indeed any algorithm for circuit minimization already implies a fast algorithm for MCSP. Thus making it easier to show $\NP$-hardness results for circuit minimization. 

An interesting distinction to note is that circuit minimization in the Boolean world is a \(\Sigma_2\)-complete problem. However, in the algebraic world, it lies within the class MA, due to the application of Polynomial Identity Testing (PIT). Specifically, given a candidate algebraic circuit, we can efficiently verify its correctness in randomized polynomial time. See \cite{BDSS24} for a more elaborate discussion on this.

Researchers in meta-complexity and algebraic complexity have been exploring the algebraic MCSP for some time now. Despite these efforts, not much is known, although some preliminary observations have been made in recent works by Baraskar et al. \cite{BDSS24} and Belova et al. \cite{BGKMS23}.

On the hardness of algebraic MCSP for special subclasses, one can obtain NP-hardness results for set-multilinear depth-3 circuits and depth-3 powering circuits essentially for free. This is a direct consequence of the hardness of computing tensor rank (and Waring rank) \cite{H90, S16}, along with standard connections between tensor and set-multilinear depth-3 circuits and Waring rank with depth-3 powering circuits \cite{BSV21}.

Recently, a beautiful work by Baraskar et al. \cite{BDSS24} showed circuit minimization for the orbit of sparse polynomials, which is a special subclass of depth-3 circuits. They established the hardness of this problem via a natural reduction from SAT. However, their work only provides hardness results for circuit minimization within this class and fails to imply the hardness of Algebraic MCSP. A reason for this is that the reduction in the work by \cite{BDSS24} also genuinely requires high-degree polynomials, making all dense representations very verbose and insufficient to imply the hardness of MCSP.

As a direct consequence of the $\NP$-hardness of order finding for ROABPs, we can deduce hardness for circuit minimization and Algebraic MCSP for ROABPs. This is because, contrary to the hard instance of \cite{BDSS24}, our hardness can arise even from constant-degree polynomials. We state these observations formally below.

\begin{corollary}[Circuit Minimization / Algebraic MCSP for ROABPs is $\NP$-hard]\label{cor:ROABP-MCSP-NP-hard}

The following are true over any field $\F$.
\begin{itemize}
    \item The algebraic circuit minimization problem for ROABPs is $\NP$-hard. That is, given an \( n \)-variate polynomial \( f \) of degree \( d \), a circuit of size \( s' \) computing \( f \), and a parameter \( w \in \mathbb{N} \), deciding whether there exists an ROABP \( C \) of width at most \( w \) that computes \( f \) is $\NP$-hard.
    \item For a fixed \( d \in \mathbb{N} \), algebraic MCSP for ROABPs is $\NP$-hard. That is, given an \( n \)-variate polynomial \( f \) of degree \( d \) along with its coefficient vector of \( \binom{n+d}{d} \) field coefficients and a parameter \( w \in \mathbb{N} \), deciding whether there exists an ROABP \( C \) in \( \mathcal{C} \) of width at most \( w \) that computes \( f \) is $\NP$-hard.
\end{itemize}
\end{corollary}

\begin{proof} 
   Note that the existence of an ROABP of width less than \( w \) directly implies that there exists an order under which the ROABP width is less than \( w \). Thus, by Theorem \ref{thm:NP-hardness-worst-case-ckt} and Theorem \ref{thm:NP-hardness-worst-case-dense}, we obtain $\NP$-hardness for ROABP circuit minimization and ROABP MCSP, respectively.
\end{proof}


\section{Algorithm for order finding}\label{sec:order-finding-algo}

We now describe a procedure (\autoref{alg:order-finding}) that solves the order finding problem when given an $n$-variate, individual degree $d$ polynomial $f(\vecx)$ as a ``black-box'' (algorithm can query $f$ at any set of points).
The algorithm solves a slightly different problem, in that it is given the width $w$ as a parameter.
It should be noted that this is essentially the same problem for the case of ROABPs, since given a purported order $\sigma$ and a black-box for $f$, there is a randomized polynomial time algorithm for reconstructing an ROABP for $f$ in the order $\sigma$ (see \autoref{sec:related-works}).
Therefore, if $w$ is not provided as an input, we could just guess a $w$ and run our algorithm. If it fails or output an incorrect order, we infer that our guess for $w$ was wrong.
This gives a straight-forward way to perform a \emph{binary search} for the correct $w$.

Our algorithm runs in two stages.
In the first stage, it populates the list of all sets $T$ such that $\nisan{T}{\overline{T}}{f} \leq w$, and in the second stage, it views this list as a graph (i.e. a subgraph of the \emph{Hasse diagram} of the boolean hypercube) to then find a path in it from $\emptyset$ to $[n]$ which reads out an order $\tau$ in which the input polynomial has a width-$w$ ROABP.
As the second stage is just a standard DFS, we provide an explicit algorithm for the first stage in \autoref{alg:populate-graph}, and only outline the other parts.

The algorithm runs in deterministic time $2^{O(n)} \cdot \poly(d,w)$ in the worst case, which is consistent with the order finding problem being $\NP$-hard for some settings of the parameters, as shown in \autoref{sec:worst-case-complexity}.
In \autoref{sec:generic-and-random-case}, we show that \autoref{alg:order-finding} runs in randomized time $n^{O(\log_{(d+1)}w)} \cdot \poly(d,w)$ with high probability when the input is a random polynomial with an ROABP of width $w$ in some order, as in \autoref{defn:random-ROABP}.

\begin{algorithm}
  \caption{\textsc{PopulateGraph}}
  \label{alg:populate-graph}

  \DontPrintSemicolon
  \SetKwInOut{Input}{Input}\SetKwInOut{Output}{Output}
  \Input{~Parameters $n,d,w$ and the polynomial $f(\vecx) \sim \mathcal{P}_{S,n,d,w,\sigma}$ given as a blackbox.}
  \Output{~Families $L_1,\ldots,L_{n-1}$ of subsets $T$ with $\nisan{T}{\overline{T}}{f} \leq w$, categorized by size.}

  \BlankLine

  Set $k \leftarrow 1$ \tcp*[r]{Size of the subsets to be tested}
  Set $L_0 \leftarrow \set{\emptyset}$ \tcp*[r]{List of ``good'' subsets of size $0$}
  Set $\mathcal{C} \leftarrow \emptyset$ \tcp*[r]{Initialize the collection to be finally output}
  
  \BlankLine

  \While{$k < n$}{
    \BlankLine
    $L_k \leftarrow \emptyset$ \tcp*[r]{Initialize the list of subsets of size $k$}
    \BlankLine
    \tcc{Try and extend every subset from $L_{k-1}$}
    \For{$T_0 \in L_{k-1}$}{
        \For{$i \in [n]$}{
            Set $T \leftarrow T_0$\;
            \If{$i \not\in T$}{
                Set $T \leftarrow T \cup \set{i}$\;
                \tcc{We check the Nisan rank using \autoref{lem:nisan-rank-estimation-from-PIT}}
                \tcc{The PIT uses the same set $S$ as in $\mathcal{P}_{S,n,d,w,\sigma}$}
                \If{$T \not\in L_k$ and $\nisan{T}{\overline{T}}{f} \leq w$}{
                    Set $L_k \leftarrow L_k \cup \set{T}$\;
                }
            }
        }
    }
    Set $\mathcal{C} \leftarrow \mathcal{C} \cup \set{L_k}$\;
    Set $k \leftarrow k+1$\;
  }
  \Return{$\mathcal{C}$}
\end{algorithm}

\begin{algorithm}
  \caption{\textsc{FindOrder}}
  \label{alg:order-finding}

  \DontPrintSemicolon
  \SetKwInOut{Input}{Input}\SetKwInOut{Output}{Output}
  \Input{~Parameters $n,d,w$ and the polynomial $f(\vecx) \sim \mathcal{P}_{S,n,d,w,\sigma}$ given as a blackbox.}
  \Output{~Permutation $\tau$ such that $f$ has an ROABP of width $w$ in the order $\tau$ (w.h.p.).}

  \BlankLine

  Let $\mathcal{C} = $ \texttt{PopulateGraph}($n,d,w,f$)\;
  Generate a graph $G$ using $\mathcal{C}$\;
  \tcc{Vertex set of $G$ is the union of $L_0,L_1,\ldots,L_n$.}
  \tcc{There is an edge from $T$ to $T'$, if $T' = T \cup \set{t}$ for some $t \in [n] \setminus T$.}
  \BlankLine
  Let $(\emptyset,T_1,T_2,\ldots,T_{n-1},[n])$ be a path in $G$\tcp*[r]{Found using, say a DFS of $G$}
  Set $\tau$: $\tau(i) = t$ if and only if $T_i = T_{i-1} \cup {t}$, for each $i \in [n]$\;
  \BlankLine
  \Return{$\tau$}
\end{algorithm}

\section{Complexity in the generic and average case}\label{sec:generic-and-random-case}

We now turn to showing that ``on average'', \autoref{alg:order-finding} runs in time that is much better than $2^{n}$ with high probability.
In particular, it runs in randomized time $n^{O(\log_{d+1} w)} \poly(d,w)$ for $n$-variate, individual degree $d$ polynomials that have ROABPs of width $w$; this running time is polynomial when $w = \poly(d)$, and quasi-polynomial even when $d$ is a constant and $w = \poly(n)$.

We start by outlining the key ideas using the language of generic ROABPs, and then extend the same arguments to the random case (see \autoref{subsec:generic-and-random-ROABPs} for the formal definitions).

\subsection{The generic case}\label{subsec:complexity-generic-case}

The simple intuition here is as follows.
We expect the order-finding algorithm to work better in the generic case, because for any order $\sigma$, a generic polynomial that has a ``small'' ROABP in that order does not satisfy any additional conditions, essentially by definition. 
What then remains to check is that having a low-rank Nisan matrix $M^{T}(f)$ for a set $T$ that is ``not consistent with $\sigma$'' is indeed an additional, or rather an independent, condition. 
We prove this in \autoref{lem:generic-ROABPs-bad-partitions}.

We use some elementary concepts from algebraic geometry (e.g. algebraic varieties) in this subsection.
While the intuitive meaning of these concepts should be clear from the discussion below, we point the reader to any text on algebraic geometry (e.g. \cite{CLO07}) for the formal definitions.

\begin{definition}[Varieties from Nisan matrices]\label{defn:varieties-of-small-nisan-rank}
    Let $\F$ be the field of rational, real or complex numbers.
    For any $n,d,w \in \N$ and $T \subseteq [n]$, we define for $N := (d+1)^n$, the set $V_{w,T} \subseteq \F^{N}$ to be the smallest algebraic variety that contains the coefficient vectors of all $n$-variate polynomials individual degree $d$ that satisfy $\nisan{T}{\overline{T}}{f} \leq w$.

    Note that $V_{w,T}$ is generated by the $(w+1) \times (w+1)$ minors of the Nisan matrix $M^{T}()$. That is, every point in $f \in V_{w,T}$ is a common zero of all those minors (which are polynomials in coefficients of $f$).
\end{definition}

Analogously, we define the variety of polynomials that have ROABPs of some width in some fixed order. This is precisely the variety of ``generic ROABPs''.
\begin{definition}[Generic polynomials with ROABPs]\label{defn:generic-polynomials-with-ROABPs}
    Let $\F$ be the field of rational, real or complex numbers.
    For any $n,d,w \in \N$ and $\sigma \in s_n$, we define for $N := (d+1)^n$, the set $V_{w,\sigma} \subseteq \F^{N}$ to be the smallest algebraic variety that contains the coefficient vectors of all $n$-variate polynomials individual degree $d$ that have an ROABP of width $w$ in the order $\sigma$.
\end{definition}

The following relation between $V_{w,\sigma}$ and the appropriate $V_{w,T}$s is an immediate consequence of Nisan's characterization (\autoref{thm:Nisan-characterization}). Note that $V_{w,\emptyset}$ and $V_{w,[n]}$ are trivial varieties.
\begin{observationwp}\label{obs:ROABP-variety-and-Nisan-rank-variety}
    For any $w \in \N, \sigma \in s_n$, let $T_1, T_2, \ldots, T_{n-1}, T_n = [n]$ be the prefixes of $\sigma$ as defined in \autoref{defn:prefix-of-permutation}.
    Then, $V_{w,\sigma} = V_{w,T_1} \cap V_{w,T_2} \cap \cdots \cap V_{w,T_{n-1}}$.
\end{observationwp}

In the language of the varieties defined above, the statement ``having a small rank under a partition that is not consistent with $\sigma$ is an additional condition'' is formalized as follows.
\begin{lemma}\label{lem:generic-ROABPs-bad-partitions}
    Let $\sigma \in s_n$ be an arbitrary permutation, and let $n,d,w \in \N$ be any integers satisfying $n \geq 3 \cdot \log_{(d+1)}w$.
    
    Then for any size $r$ such that $k \leq r \leq  n - k$, where $k := \floor{\log_{(d+1)}w} + 1$, and for any set $T$ such that neither $T$ nor $[n] \setminus T$ is a prefix of $\sigma$, we have that $V_{w,\sigma} \cap V_{w,T} \subsetneq V_{w,\sigma}$.
\end{lemma}

We will prove the lemma (at the end of \autoref{subsec:constructing-hard-polys}) by showing the existence of a polynomial $f(\vecx)$ that lies in $V_{w,\sigma}$, but not in $V_{w,T}$. In other words, we construct (for every valid choice of the parameters) a polynomial $f$ that has a width-$w$ ROABP in the order $\sigma$ but has $\nisan{T}{\overline{T}}{f} > w$.
We do this in two steps.
First, we construct such a polynomial for $\sigma = \idperm$ and for some structured set $T$ in \autoref{lem:bad-poly-for-fixed-set}. We then show how the same construction can be generalized to all sets of appropriate sizes, and all orders in \autoref{lem:bad-polys-full-generality}.

Using \autoref{lem:generic-ROABPs-bad-partitions}, we can now say that the varieties corresponding to the ``inconsistent partitions'' do not cover the variety $V_{w,\sigma}$.
\begin{lemma}\label{lem:bad-partition-varieties-dont-cover}
    Let $\F$ be $\Q$, $\R$ or $\C$.
    Let $\sigma \in s_n$ be an arbitrary permutation, and let $n,d,w \in \N$ be any integers satisfying $n \geq 3 \cdot \log_{(d+1)}w$.
    
    Suppose $\mathcal{T}$ is the family of all sets $T$ such that neither $T$ nor $[n] \setminus T$ is a prefix of $\sigma$, and $k \leq \abs{T} \leq  n - k$, where $k := \floor{\log_{(d+1)}w} + 1$.
    Then $V_{w,\sigma} \not\subset \bigcup_{T \in \mathcal{T}} V_{w,T}$.
\end{lemma}
\begin{proof}
    Using the definition of a random ROABP (\autoref{defn:random-ROABP}), notice that the coefficient vectors in $V_{w,\sigma}$ are polynomials in the entries of the coefficient matrices.
    This means that there is a polynomial map $P : \F^M \rightarrow \F^N$, for $M = n(d+1)w^2$ and $N = (d+1)^n$, which maps the entries of the coefficient matrices to coefficient vectors inside $V_{w,\sigma}$.
    Since $\F^M$ is an irreducible\footnote{A variety is irreducible if it cannot be written as a union of two strictly smaller varieties.} variety, so is $V_{w,\sigma}$.

    Therefore, \autoref{lem:generic-ROABPs-bad-partitions} tells us that the dimension of $V_{w,\sigma} \cap V_{w,T}$ is strictly smaller than that of $V_{w,\sigma}$.
    Finally, since $\mathcal{T}$ is a finite collection (and $\F$ is infinite), we cannot cover $V_{w,\sigma}$ using finitely many varieties of strictly smaller dimensions.
\end{proof}

\begin{theorem}[Finding an optimal order in the generic case]\label{thm:generic-case-runtime}
    Let $\F$ be the field of rational, real or complex numbers.
    Let $\sigma \in s_n$ be an arbitrary permutation, and let $n,d$ and $w$ be any positive integers satisfying $n \geq 3 \cdot \log_{(d+1)}w$.

    Then there is a randomized algorithm $R$ that, when given a generic $n$-variate polynomial $f(\vecx)$ of individual degree $d$ that has an ROABP of width $w$ in the order $\sigma$ as a \emph{black-box}, outputs some $\tau \in s_n$, such that $f$ has an ROABP of width at most $w$ in the order $\tau$.

    Further, with high probability, the running time of $R$ is $n^{O(\log_{(d+1)}w)} \cdot \poly(d,w)$.
\end{theorem}
\begin{proof}
    The algorithm $R$ is exactly\footnote{The only tiny change is that since we do not have a predefined set $S \subseteq \F$, the rank-estimation has to choose such a set, which is taken to be $\set{1,\ldots,2^{2n+1}ndw^2}$.} the order finding algorithm outlined in \autoref{alg:order-finding}.

    By \autoref{lem:bad-partition-varieties-dont-cover}, a generic polynomial exhibits Nisan matrices of rank strictly more than $w$ for any $T$ that is ``not consistent'' with the order $\sigma$.
    Thus, the algorithm \texttt{PopulateGraph} (\autoref{alg:populate-graph}) generates lists $L_1,\ldots,L_{n-1}$, such that for $\log_{(d+1)}w < k < n - \log_{(d+1)}w$, $\abs{L_k} \leq 2$.
    
    Next, note that \autoref{alg:populate-graph} performs an identity test as described in \autoref{lem:nisan-rank-estimation-from-PIT} at most $n \cdot \abs{L_k}$ times, for each $0 \leq k \leq n-1$. This is because \autoref{alg:populate-graph} only attempts to extend the sets that the previous iteration has generated.
    Therefore, with probability at least $\inparen{1 - 2^{-n}}$, the algorithm performs at most $\insquare{2 (n - 2 \log_{(d+1)}w) + 2 \cdot \log_{(d+1)}w \cdot \binom{n}{\log_{(d+1)}w}} < 2^n$ identity tests, each of which fails with probability at most $\sfrac{nd(w+1)^2}{\abs{S}} \leq 2^{-2n}$, when $S$ is chosen to be appropriately large.
    Thus, the probability that all these PITs succeed, is at least $(1 - 2^n 2^{-2n})$, by the union bound. 
    Therefore, $R$ runs in time $n^{O(\log_{(d+1)}w)}$ and outputs a correct order with probability at least $(1 - 2^{-n})^2 \geq 1 - 2^{n-1}$.
\end{proof}

We remark that since $M^{T}(f)$ has $(d+1)^{\abs{T}}$ rows, $\nisan{T}{\overline{T}}{f} \leq (d+1)^{\abs{T}}$ for any $f$.
Thus, any set $T$ of size at most $k = \log_{d+1} w$ results in a Nisan matrix of rank at most $w$, and therefore, the factor of $n^{O(\log_{(d+1)}w)}$ is unavoidable in the running time for \autoref{alg:populate-graph}.

\subsection{Constructing the polynomials}\label{subsec:constructing-hard-polys}

\begin{lemma}\label{lem:bad-poly-for-fixed-set}
    Let $\F$ be any field, and $n,d$ and $w$ be any positive integers satisfying $n \geq 3 \cdot \log_{(d+1)}w$.
    Suppose $T \in [n]$ satisfies the following three properties.
    \begin{enumerate}\itemsep0pt
        \item $T$ has a reasonable size: $\floor{\log_{(d+1)}w} + 1 \leq \abs{T} \leq n/2$.
        \item $T$ is ``left-heavy'': $\abs{T \cap \set{1,\ldots,\floor{n/2}}} \geq \abs{T \cap \set{\floor{n/2} + 1,\ldots,n}}$.
        \item $T$ is not a prefix with respect to $\idperm$: $T \neq \set{1,\ldots,\abs{T}}$.
    \end{enumerate}
    Then there exists an $n$-variate, individual-degree $d$ polynomial $f(\vecx) \in \F[\vecx]$ that has a width $w$ ROABP in the order $\idperm$, but satisfies $\nisan{T}{\overline{T}}{f} > w$.
\end{lemma}
\begin{proof}
    Let $k = \floor{\log_{(d+1)}w} + 1$, so that $k \leq \abs{T} \leq n/2$. We will define a polynomial $f$ that depends only on $2k$ of the $n$ variables, split into two sets: $L$ and $R$ of size $k$ each.
    For $L = \set{i_1,i_2,\ldots,i_k}$ and $R = \set{j_1,j_2,\ldots,j_k}$, we will then define
    \[
        f(\vecx) = \prod_{\ell \in [k]} \inparen{ 1 + x_{i_{\ell}} x_{j_{\ell}} +  x_{i_{\ell}}^2 x_{j_{\ell}}^2 + \cdots +  x_{i_{\ell}}^d x_{j_{\ell}}^d }
    \]
    The idea here is to choose $L$ and $R$ in a way that ensures that: first, $L \subseteq T$ and $R \subseteq [n] \setminus T$, and second, any prefix of $\idperm$ (or its complement) fully contains some pair $(i_{\ell},j_{\ell})$.
    These conditions would then ensure that $\nisan{T}{\overline{T}}{f} = (d+1)^{k} = (d+1)^{\floor{\log_{(d+1)}w} + 1} > w$, and that for all $i \in [n]$, $\nisan{[i]}{[n] \setminus [i]}{f} \leq (d+1)^{k-1} \leq w$, as required (see \autoref{obs:properties-nisan-matrices}).

    \begin{figure}
        \centering
        \begin{tikzpicture}

            \begin{scope}[shift={(-4.3,0)}]
                \draw (-3.2,0) -- (-2,0) (-1,0) -- (-0.7,0) (0.7,0) -- (1,0) (2,0) -- (3.2,0);
                \draw[dotted] (-3.5,0) -- (-3.2,0) (-2,0) -- (-1,0) (-0.7,0) -- (0.7,0) (1,0) -- (2,0) (3.2,0) -- (3.5,0);
                \fill (-3,0) circle (1.5pt) (-2,0) circle (1.5pt) (-1,0) circle (1.5pt) (0,0) circle (1.5pt);
                \fill (1,0) circle (1.5pt) (2,0) circle (1.5pt) (3,0) circle (1.5pt);
                
                \node (one) at (-3.6,-0.5) {$1$};
                \node (n) at (3.6,-0.5) {$n$};
                \fill (-3.6,0) circle (1.5pt) (3.6,0) circle (1.5pt);
                
                \draw [color=NavyBlue] (-2,0) to[out=85, in=95] (-1,0);
                \draw [color=NavyBlue] (1,0) to[out=85, in=95] (2,0);
                \node (i11) at (-2,-0.5) {$i_1$};
                \node (j11) at (-1,-0.5) {$j_1$};
                \node (i21) at (1,-0.5) {$i_2$};
                \node (j21) at (2,-0.5) {$j_2$};
            \end{scope}

            \begin{scope}[shift={(4.3,0)}]
                \draw (-3.2,0) -- (-2,0) (-1,0) -- (0,0) (2,0) -- (3.2,0);
                \draw[dotted] (-3.5,0) -- (-3.2,0) (-2,0) -- (-1,0) (0,0) -- (2,0) (3.2,0) -- (3.5,0);
                \fill (-3,0) circle (1.5pt) (-2,0) circle (1.5pt) (-1,0) circle (1.5pt) (0,0) circle (1.5pt);
                \fill (2,0) circle (1.5pt) (3,0) circle (1.5pt);
                
                \node (one) at (-3.6,-0.5) {$1$};
                \node (n) at (3.6,-0.5) {$n$};
                \fill (-3.6,0) circle (1.5pt) (3.6,0) circle (1.5pt);
                
                \draw [color=NavyBlue] (-2,0) to[out=85, in=95] (-1,0);
                \draw [color=NavyBlue] (0,0) to[out=80, in=100] (2,0);
                \node (j12) at (-2,-0.5) {$j_1$};
                \node (i12) at (-1,-0.5) {$i_1$};
                \node (i22) at (0,-0.5) {$i_2$};
                \node (j22) at (2,-0.5) {$j_2$};
            \end{scope}

            \begin{scope}[shift={(0,-2.3)}]
                \draw (-3.2,0) -- (-1.7,0) (-0.3,0) -- (3.2,0);
                \draw[dotted] (-3.7,0) -- (-3.2,0) (-2,0) -- (0,0) (3.2,0) -- (3.7,0);
                \fill (-3,0) circle (1.5pt) (-2,0) circle (1.5pt) (-1,0) circle (1.5pt) (0,0) circle (1.5pt);
                \fill (1,0) circle (1.5pt) (2,0) circle (1.5pt) (3,0) circle (1.5pt);
                
                \node (one) at (-3.8,-0.5) {$1$};
                \node (n) at (3.8,-0.5) {$n$};
                \fill (-3.8,0) circle (1.5pt) (3.8,0) circle (1.5pt);
                
                \draw [color=BrickRed] (-2,0) to[out=75, in=105] (1,0);
                \draw [color=BrickRed] (0,0) to[out=80, in=100] (2,0);
                \node (i13) at (-2,-0.5) {$i_1$};
                \node (i23) at (0,-0.5) {$i_2$};
                \node (j13) at (1,-0.5) {$j_1$};
                \node (j23) at (2,-0.5) {$j_2$};
            \end{scope}
            
        \end{tikzpicture}
        \caption{Examples of \textcolor{NavyBlue}{good} (top) and \textcolor{BrickRed}{bad} (bottom) choices for $(i_1,j_1)$ and $(i_2,j_2)$.}
        \label{fig:choosing-first-two-pairs}
    \end{figure}
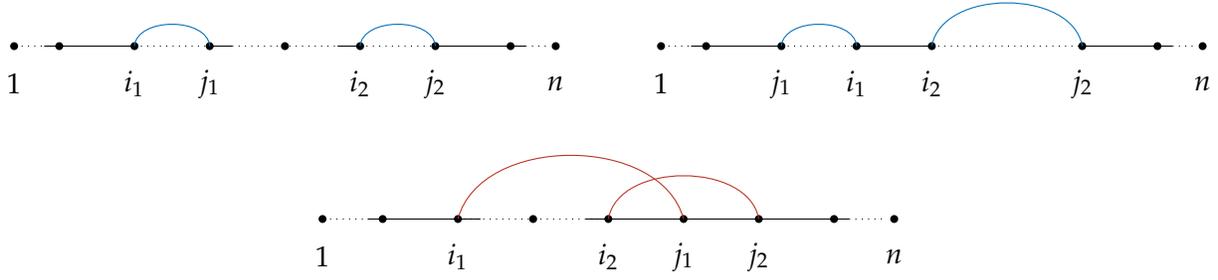

    We now describe how we can choose such sets $L$ and $R$.

    Observe that if $(i_1,j_1)$ and $(i_2,j_2)$ are such that the first pair is completely ``on the left'' of the second (that is $\max(i_1,j_1) < \min(i_2,j_2)$, see \autoref{fig:choosing-first-two-pairs}), then we would ensure the second condition.
    Therefore, it is enough to show that such indices $i_1,j_1$ and $i_2,j_2$ exist with respect to any set $T$ that satisfies the three properties in the hypothesis.

    We choose $i_1$ to be the smallest index in $T$ and $j_1$ to be the smallest index that is outside $T$. Similarly, we choose $i_2$ to be the largest index in $T$ and $j_2$ to be the largest index outside $T$.
    
    Since $i_1 < i_2$ and $j_1 < j_2$ by definition, we only need to show that $i_1 < j_2$ and $j_1 < i_2$.
    As $T$ is left heavy, $i_1 < n/2$ and further since $\abs{T} \leq n/2$, $j_2 > n/2$, thus $i_1 < j_2$.
    Also, $j_1 > i_2$ is possible only if $T = \set{1,\ldots,\abs{T}}$, which contradicts $T$ not being a prefix with respect to $\idperm$.

    Finally, we choose $i_3,\ldots,i_k$ arbitrarily from $T$, and $j_3,\ldots,j_k$ arbitrarily from $[n] \setminus T$.
    This is fine, since our choice of the first two pairs already ensures that $\nisan{[i]}{[n] \setminus [i]}{f} \leq w$ for any $i \in [n]$, and since $T$ contains exactly half of all the $k$ pairs being chosen, $\nisan{T}{\overline{T}}{f} = (d+1)^k > w$.
    This finishes the proof using Nisan's characterization (\autoref{thm:Nisan-characterization}).
\end{proof}

\begin{remark}\label{rmk:about-left-heaviness}
    Actually, the left-heaviness of the set $T$ can be relaxed. That is, if $T$ is right-heavy instead, then we would ``flip'' the choice of $i_1,j_1,i_2$ and $j_2$ ``about the middle'', by choosing $i_1,j_1$ to be the largest indices inside and outside $T$ and $i_2,j_2$ to be the smallest indices inside and outside $T$.
    One just needs to ``reflect the argument'' about $n/2$: the first pair is to the right of the second pair and so on, for the proof to work.
    We skip the details to keep the proof simpler.
\end{remark}

Next, we extend the above result to all sets $T$ that do not trivially lead to $\nisan{T}{\overline{T}}{f} \leq w$ when the order is $\idperm$.
This essentially boils down to working with either $T$ or $\overline{T} = [n] \setminus T$, and with $\id$ or $\rev(\id)$, to ensure that the hypothesis of \autoref{lem:bad-poly-for-fixed-set}.
This extension then lets us bound the probability that a random polynomial with an ROABP in the order $\idperm$ has small Nisan-rank for an ``incorrect'' set $T$, as follows.

\begin{lemma}\label{lem:bad-polys-full-generality}
    Let $n,d$ and $w$ be any positive integers satisfying $n \geq 3 \cdot \log_{(d+1)}w$, and let $\sigma \in s_n$ be arbitrary.
    Suppose $T \subseteq [n]$ is such that $k \leq \abs{T} \leq  n - k$ for $k := \floor{\log_{(d+1)}w} + 1$, and neither $T$ nor $([n] \setminus T)$ is a prefix with respect to $\sigma$.
    
    Then there exists an $n$-variate, individual-degree $d$ polynomial $f(\vecx)$ that has a width $w$ ROABP in the order $\sigma$, but satisfies $\nisan{T}{\overline{T}}{f} > w$.
\end{lemma}
\begin{proof}
    We can assume $\sigma = \idperm$ by just renaming the variables in $f$ and the indices in $T$ to adjust for $\sigma$. 
    That is, for $\tau = \sigma^{-1}$, if $T$ satisfies the hypothesis of the lemma for $\sigma$, then $\tau(T) = \set{\tau(t) : t \in T}$ satisfies the hypothesis for $\idperm$, and if $f(x_1,\ldots,x_n)$ satisfies the conclusion for $T$, then  $f(x_{\tau(1)},\ldots,x_{\tau(n)})$ satisfies the lemma for $\tau(T)$.
    
    We now argue that we can make some simplifying assumptions on $T$ so that we can then invoke \autoref{lem:bad-poly-for-fixed-set}.
    \begin{itemize}
        \item $\abs{T} \leq n/2$: If this is not the case, then we work with the complement $\overline{T} = [n] \setminus T$, since $\nisan{T}{\overline{T}}{f} = \nisan{\overline{T}}{T}{f}$ for any polynomial $f$ and any set $T$.
        \item $\abs{T \cap \set{1,\ldots,\floor{n/2}}} \geq \abs{T \cap \set{\floor{n/2} + 1,\ldots,n}}$: If this is not the case, then we ``reverse all coordinates'' by applying the permutation $\rev(\idperm)$ and work with the reversed set $\mathrm{rev}(T) = \set{ (n+1) - i : i \in T}$.
        Again, this is without loss of generality, since for any permutation $\sigma \in s_n$, the set of polynomials that have an ROABP of width $w$ in the order $\sigma$ is exactly the same as the same set corresponding to $\rev(\sigma)$, as a direct consequence of \autoref{thm:Nisan-characterization}.
        Refer to \autoref{rmk:about-left-heaviness} for further details on this assumption.
    \end{itemize}
    Now, since $T$ (or $[n] \setminus T$) was not a prefix with respect to $\id$, we get that the simplified $T$ is not a prefix with respect to $\idperm$.
    Thus, we can apply \autoref{lem:bad-poly-for-fixed-set} to obtain an $f(\vecx) \in \F[\vecx]$ such that $\nisan{T}{\overline{T}}{f} > w$.
\end{proof}

\begin{proof}[Proof of \autoref{lem:generic-ROABPs-bad-partitions}]
    We use \autoref{lem:bad-polys-full-generality} to find a polynomial $f(\vecx)$ that has an ROABP of width $w$ in the order $\sigma$, but has $\nisan{T}{\overline{T}}{f} > w$.
    Thus, the coefficient vector of $f$ is a point in $V_{w,\sigma}$ that is not inside $V_{w,T}$, and therefore $V_{w,\sigma} \cap V_{w,T} \subsetneq V_{w,\sigma}$.
\end{proof}

\subsection{Finding order of a random ROABP}\label{subsec:complexity-random-case}

We now give the ``average case'' analogues of our results in \autoref{subsec:complexity-generic-case}. At a high level, these results are obtained by carefully applying the Schwartz-Zippel lemma (\autoref{lem:schwartz-zippel}) to bound the probabilities of a random ROABP (see \autoref{defn:random-ROABP}) satisfying any additional conditions that are not satisfied by a generic ROABP.
An upshot of this exercise is that this allows us to talk about a random ROABP that is sampled to have coefficient matrices coming from large enough finite fields.

\begin{lemma}\label{lem:UB-on-bad-cases-random-roabp}
    Let $S \subseteq \F$ be any set of size $2^{2n+1} \cdot n \cdot w$, let $\sigma \in s_n$ be an arbitrary permutation, and let $n,d$ and $w$ be any positive integers satisfying $n \geq 3 \cdot \log_{(d+1)}w$.
    
    Then for any size $r$ such that $k \leq r \leq  n - k$ for $k := \floor{\log_{(d+1)}w} + 1$, and for a set $T$ such that neither $T$ nor $[n] \setminus T$ is a prefix of $\sigma$, the probability that the random ROABP $\mathcal{P}_{S,n,d,w,\sigma}$ has $\nisan{T}{\overline{T}}{\mathcal{P}} \leq w$, is at most $2^{-2n}$.
\end{lemma}
\begin{proof}
    For any polynomial $f$, we use $\overline{\coeff}(f)$ to denote the vector of coefficients of $f$.

    First, we use \autoref{lem:bad-polys-full-generality} to find a polynomial $f(\vecx)$ that has an ROABP of width $w$ in the order $\sigma$, but has $\nisan{T}{\overline{T}}{f} > w$.
    Now let $h_0$ be a $(d+1)^n$-variate polynomial that is a $(w+1) \times (w+1)$ minor of the Nisan matrix for $T$ (viewing the coefficients of the polynomial as variables) $M^{T}$, such that $h_0(\overline{\coeff}(f)) \neq 0$.
    Such a minor, and thus the corresponding $h_0$, exists precisely because $\nisan{T}{\overline{T}}{f} > w$.

    Now consider substituting the coefficients of $\mathcal{P}_{S,n,d,w,\sigma}$: $\overline{\coeff}(\mathcal{P})$, in $h_0$.
    Since $h_0$ is a polynomial of degree $(w+1)$ in these coefficients, and the coefficients are polynomials of degree $n$ in the entries of the random matrices $\set{\mathcal{A}_{i,j}}$ (see \autoref{obs:degree-of-coeffs-of-P-in-As}), their composition is a polynomial of degree $n(w+1)$ in the entries of $\set{\mathcal{A}_{i,j}}$.
    Name this composed polynomial $g(\vec{\mathcal{A}})$.

    Note that $g(\vec{\mathcal{A}})$ is a nonzero polynomial, since there is a realization of the variables that precisely yields the width-$w$ ROABP for $f(\vecx)$ in the order $\idperm$, and the coefficient vector of $f$ is not a root of $h_0$.
    This gives us the following bound on the probability of $\nisan{T}{\overline{T}}{\mathcal{P}} \leq w$, as claimed.
    \begin{align*}
        \Pr_{\vec{\mathcal{A}}} \insquare{\nisan{T}{\overline{T}}{\mathcal{P}} \leq w} &= \Pr_{\vec{\mathcal{A}}} \insquare{ \bigwedge_{\text{$h$ is a $(w+1) \times (w+1)$ minor of $M_T$}} h(\overline{\coeff}(\mathcal{P})) = 0}\\
        \text{($h_0$ is one such minor)} &\leq \Pr_{\vec{\mathcal{A}}} \insquare{ h_0(\overline{\coeff}(\mathcal{P})) = 0}\\
        \text{(Substituting for $\overline{\coeff}(\mathcal{P})$)} &= \Pr_{\vec{\mathcal{A}}} \insquare{ g(\vec{\mathcal{A}}) = 0}\\
        \text{(Schwartz-Zippel lemma)} &\leq \frac{\deg(g)}{\abs{S}} \leq \frac{2nw}{\abs{S}} \leq 2^{-2n} \qedhere
    \end{align*}
\end{proof}

\begin{lemma}\label{lem:UB-for-algo-random-case}
    Let $S \subseteq \F$ be any set of size $\geq 2^{2n+1} \cdot n \cdot w$, let $\sigma \in s_n$ be an arbitrary permutation, and let $n,d$ and $w$ be any positive integers satisfying $n \geq 3 \cdot \log_{(d+1)}w$.

    Then for $k = \floor{\log_{(d+1)}w} + 1$, and any $r$ such that $k \leq r \leq n - k$ the probability that the list $L_r$ in \autoref{alg:populate-graph} has size greater than $2$ for a random $f(\vecx) \sim \mathcal{P}_{S,n,d,w,\sigma}(\vecx)$, is at most $2^{-n}$.
\end{lemma}
\begin{proof}
    Observe that for any fixed set $T$ of size $r$ that satisfies the hypothesis, we can just invoke \autoref{lem:UB-on-bad-cases-random-roabp} to derive that the probability that $\nisan{T}{\overline{T}}{\mathcal{P}} \leq w$ is at most $\sfrac{2nw}{\abs{\F}} = 2^{-2n}$.
    Then by the union bound, the probability that any such set $T$ exists is at most $2^n \cdot 2^{-2n} = 2^{-n}$.
    
    As $\mathcal{P}(\vecx)$ has a width-$w$ ROABP, each list $L_r$ in \autoref{alg:populate-graph} has size at least $2$, corresponding to the prefixes with respect to the order $\sigma$ and the complement of that prefix.
    If $\abs{L_r} > 2$ for the given parameters, then there exists a subset $T$ such that neither $T$ nor $[n] \setminus T$ is a prefix of $\sigma$, with $\nisan{T}{\overline{T}}{f} \leq w$.
    The probability that such a set $T$ is at most $2^{-n}$, from the argument above.
\end{proof}

\begin{theorem}[Finding an optimal order in the average case]\label{thm:random-case-runtime}
    Over any large enough field $\F$, let $S \subseteq \F$ be of size $\geq 2^{2n+1} \cdot (n d w^2)$, let $\sigma \in s_n$ be an arbitrary permutation, and let $n,d$ and $w$ be any positive integers satisfying $n \geq 3 \cdot \log_{(d+1)}w$.

    Then there is a randomized algorithm $R$ that, when given a random polynomial $f \sim \mathcal{P}_{S,n,d,w,\sigma}$ as a \emph{black-box}, outputs some $\tau \in s_n$, such that $f$ has an ROABP of width at most $w$ in the order $\tau$, with probability at least $(1 - 2^{n-1})$.

    Further, with probability at least $(1 - 2^{n-1})$, $R$ runs in time $n^{O(\log_{(d+1) w})} \cdot \poly(d,w)$.
\end{theorem}
\begin{proof}
    The algorithm $R$ is the order-finding algorithm outlined in \autoref{alg:order-finding}.
    By \autoref{lem:UB-for-algo-random-case}, with probability at least $\inparen{1 - 2^{-n}}$, the algorithm \texttt{PopulateGraph} (\autoref{alg:populate-graph}) generates lists $L_1,\ldots,L_{n-1}$, such that for $\log_{(d+1)}w < k < n - \log_{(d+1)}w$, $\abs{L_k} \leq 2$.
    
    Next, note that \autoref{alg:populate-graph} performs an identity test as described in \autoref{lem:nisan-rank-estimation-from-PIT} at most $n \cdot \abs{L_k}$ for each $0 \leq k \leq n-1$. This is because \autoref{alg:populate-graph} only attempts to extend the sets that the previous iteration has generated.
    Therefore, with probability at least $\inparen{1 - 2^{-n}}$, the algorithm performs at most $2 (n - 2 \log_{(d+1)}w) + 2 \cdot \log_{(d+1)}w \cdot \binom{n}{\log_{(d+1)}w}$ identity tests, each of which fails with probability at most $\sfrac{nd(w+1)^2}{\abs{S}} \leq 2^{-2n}$.
    Thus, the probability that all these PITs succeed, is at least $(1 - 2^n 2^{-2n})$, by the union bound, and hence $R$ runs in the promised time and outputs a correct order, with probability at least $(1 - 2^{-n})^2 \geq 1 - 2^{n-1}$, as required.
\end{proof}

\section{Approximate order finding}

As mentioned earlier, since $\cutwidth$ is known to be hard to approximate up to any constant factor assuming the Small Set Expansion (SSE) conjecture~\cite{APW12}.
However, one can still ask if the hardness of approximation for ROABP-width follows from a more widely believed assumption.

There are two relevant results in this regard.
First is a hardness of approximation (under $\P \neq \NP$) result for approximating the OBDD-size of a boolean function, due to Sieling~\cite{S02}; however this result is quite technical, and it is unclear to us if it implies such a result for ROABP-width (or size).
The other result is due to Amb\"{u}hl, Mastrolilli and Svensson~\cite{AMS11}, which rules out a PTAS for the \emph{minimum/optimal linear arrangement} problem, under the assumption that $\NP \not\subseteq \SUBEXP$. Again, due to the exactness of our reduction from width of any cut in a linear arrangement to the width of the ROABP in the corresponding layer (see \autoref{lem:cutwidth-to-order-finding}), this result directly rules out a PTAS for ROABP-size.

Turning to the problem of finding an order that approximates the optimal ROABP-width, we (the authors) do not know of a hardness result that rules out a constant approximation algorithm, or even a PTAS, for the $\cutwidth$ problem, under any stronger assumption that the SSE conjecture.

\subsection{Constant-factor Approximation implies a PTAS}

We show that for any constant $\alpha$, an $\alpha$-approximation algorithm for $\SearchCktROwidth$ implies a PTAS for it, which is formally stated below. Thus, combined with our reduction from $\cutwidth$ to ROABP-width, any result that rules out a PTAS for $\cutwidth$ under any strong assumption(s), would also rule out any constant-factor approximation algorithm for ROABP-width.

\begin{theorem}\label{thm:constant-approx-to-PTAS-ckt}
    Suppose for a constant $\alpha > 1$ that $\mathtt{ApproxWidth}$ is a deterministic polynomial time algorithm, which when given the tuple $(C,n,d)$ --- where $C$ is an algebraic circuit computing an $n$-variate polynomial $f(\vecx)$ of individual degree $d$ --- outputs an order $\sigma$ such that $f$ has an ROABP of width $w^{*}$ in the order $\sigma$, and $w^{*}$ is at most $\alpha$ times the optimal ROABP-width of $f(\vecx)$.

    Then, there is a deterministic algorithm $\mathtt{WidthPTAS}$, that takes as input $(C,n,d,\epsilon)$ (where $C,n,d$ are as above), and outputs an order $\sigma$ such that the ROABP-width of the polynomial computed by $f$ in the order $\sigma$, is at most $(1 + \epsilon)$ times its optimal ROABP-width. 
    Further, this algorithm runs in time $m^{O_{\alpha}(\sfrac{1}{\epsilon})}$ on all inputs of length $m$.
\end{theorem}

We prove the following lemma, which gives a way of multiplicatively increasing the optimal ROABP-width of a polynomial with an easy transformation.
\begin{lemma}[Tensoring ROABP width]\label{lem:tensoring-roabp-width}
    Over any field $\F$, let $f(x_1,\ldots,x_n) \in \F[\vecx]$ be a polynomial with individual degree $d$, and let $\sigma \in s_n$ be arbitrary. Suppose the optimal ROABP-width for $f$ in the order $\sigma$ is $w$, then the following polynomial has width $w^k$ in the order $\sigma$, for any integer $k \geq 1$.
    \[
        g(\vecx) := f(x_1,\ldots,x_n) \cdot f(x_1^{d+1},\ldots,x_n^{d+1}) \cdot \cdots \cdot f\inparen{x_1^{(d+1)^{k-1}},\ldots,x_n^{(d+1)^{k-1}}}
    \]
\end{lemma}

We introduce below the concept of a lifted partition to prove the theorem.
\begin{definition}[Lifted partitions]\label{defn:lifted-partitions}
    Let $\mathbf{Y} = Y_1 \cup Y_2 \cup \cdots \cup Y_n$ be a set of variables containing $n$ \emph{blocks}: $Y_i = \set{y_{i,1},\ldots,y_{i,k}}$ for each $i \in [n]$.
    We call a partition $\mathbf{Y} = S \sqcup T$ \emph{lifted}, if for each $Y_i$, either $Y_i \subseteq S$ or $Y_i \subseteq T$.

    In other words, a \emph{lifted} partition $S \sqcup T$ has an underlying partition $A \sqcup B = [n]$ such that $S$ and $T$ are exactly the unions of the blocks in $A$ and $B$ respectively.
\end{definition}

Along the lines on \autoref{obs:properties-nisan-matrices}, lifted partitions let us increase the optimal ROABP-width multiplicatively.
\begin{lemma}\label{lem:tensoring-with-blocks}
    For a polynomial $f(x_1,\ldots,x_n)$ and any integer $k \geq 1$, let $h_k(\mathbf{Y})$ be
    \[
        h_k(\mathbf{Y}) := f(y_{1,1},\ldots,y_{n,1}) \cdot f(y_{1,2},\ldots,y_{n,2}) \cdots f(y_{1,k},\ldots,y_{n,k}),
    \]
    with $\mathbf{Y} = Y_1 \cup Y_2 \cup \cdots \cup Y_n$, and each $Y_i = \set{y_{i,1},\ldots,y_{i,k}}$.

    Suppose $\mathbf{X} = S \sqcup T$ is a \emph{lifted partition}, where the underlying partition is $[n] = A \sqcup B$.
    Then, if the $\nisan{A}{B}{f} = w$, then $\nisan{S}{T}{h_k} = w^k$.
\end{lemma}
\begin{proof}
    Let $M = M^{(A,B)}(f)$, then we show that $M^{(S,T)}(h_k)$ is precisely the Kronecker (or ``tensor'') product $M^{\otimes k} = M \otimes M \otimes \cdots \otimes M$.

    To see this, let $m = \prod_{i \in A} (\prod_{\ell \in [k]} y_{i,\ell}^{a_{i,\ell}})$ be a monomial over the variables in the set $S$: $\bigcup_{i \in A} Y_i$, and similarly, let $m' = \prod_{j \in B} (\prod_{\ell \in [k]} y_{j,\ell}^{b_{j,\ell}})$ be a monomial over the variables in the set $T$.
    Then, by definition of $h_k$, $\coeff_{h_k}(m \cdot m') = \prod_{\ell \in [k]} \coeff_{f}\inparen{\prod_{i \in A} y_{i,\ell}^{a_{i,\ell}} \cdot \prod_{j \in B} y_{j,\ell}^{b_{j,\ell}}}$.
    This essentially means that $M^{(S,T)}(h_k)[m,m'] = \prod_{\ell \in [k]} M^{(A,B)}(f)[m_{\ell},m'_{\ell}]$, where $m_{\ell} = \prod_{i \in A} y_{i,\ell}^{a_{i,\ell}}$ and $m'_{\ell}$ is defined analogously.
    This is exactly the same as saying $M^{(S,T)}(h_k) = (M^{(A,B)}(f))^{\otimes k}$. This finishes the proof.
\end{proof}

\begin{proof}[Proof of \autoref{lem:tensoring-roabp-width}]
    First, define the polynomial $h_k$ from $f$, in a similar manner as in \autoref{lem:tensoring-with-blocks}.
    Therefore, for the unique $\F$-linear map $\mu : \F[\mathbf{Y}] \rightarrow \F[\vecx]$ that satisfies $\mu : y_{i,j} \mapsto x_i^{(d+1)^{j-1}}$ for all $y_{i,j}$,  we have that $g(\vecx) = \mu(h_k)$.
    
    Thus, since the individual degree of $h_k$ is $d$, for any monomial $m_0 = \prod_{i \in [n]} \prod_{j \in [k]} y_{i,j}^{e_{i,j}}$, there is a \emph{unique} corresponding monomial $m = \prod_{i \in [n]} \prod_{j \in [k]} x_{i}^{e_{i,j} \cdot (d+1)^{j-1}}$, such that $\coeff_{g}(m) = \coeff_{h_k}(m_0)$.
    Therefore, the map $\mu$ is a bijection between the monomials of $h_k$ and $g$.
    This means that for any partition $[n] = A \sqcup B$, the matrices $M^{(A,B)}(g)$ and $M^{(S,T)}(h_k)$ are exactly the same, up to renaming their rows and columns, under the corresponding \emph{lifted partition} $(S,T)$. 
    So the $\mathrm{rank}(M^{(A,B)}(g))$ is exactly the $k$-th power of the rank of the Nisan matrix of $f$ under the partition $(A,B)$.

    Since the rank of every Nisan matrix has been raised to the power $k$ in the transformation from $f$ to $g$, the width of an optimal ROABP for $g$ under any order $\sigma$ is exactly the $k$-th power of that for $f$, using Nisan's characterization (\autoref{thm:Nisan-characterization}).
\end{proof}

\begin{proof}[Proof of \autoref{thm:constant-approx-to-PTAS-ckt}]
    We first describe the algorithm $\mathtt{WidthPTAS}$ and then argue its correctness and running time.
    Suppose the input is $(C,n,d,\epsilon)$, where $C$ computes the polynomial $f(x_1,\ldots,x_n)$.
    \begin{enumerate}
        \item Let $k = \ceil{\sfrac{\log \alpha}{\log (1+\epsilon)}}$.
        \item Define $\tilde{C}(\vecx) := C(x_1,\ldots,x_n) \cdot C\inparen{x_1^{(d+1)},\ldots,x_n^{(d+1)}} \cdots C\inparen{x_1^{(d+1)^{k-1}},\ldots,x_n^{(d+1)^{k-1}}}$.
        \item Let $w^* = \mathtt{ApproxWidth}(\tilde{C},n,d_k)$ for $d_k = (d+1)^{k} - 1$.
        \item Compute $\hat{w} = (w^*)^{1/k}$, and return $\hat{w}$.
    \end{enumerate}

    \paragraph*{Correctness.} Suppose the optimal ROABP width of the input $f(\vecx)$ is $w$.
    From \autoref{lem:tensoring-roabp-width}, it is clear that the polynomial computed by $\tilde{C}$ has ROABP-width exactly $w^k$, whenever $f(\vecx)$ has width $w$.
    Since we are assured that $w^{*} \leq \alpha \cdot w^k$, $\hat{w} = (w^*)^{1/k} \leq \alpha^{1/k} \cdot w$.
    Therefore, $\hat{w} \leq \alpha^{\sfrac{\log (1+\epsilon)}{\log \alpha}} \cdot w$, which is at most $(1+\epsilon) w$, as required.

    \paragraph*{Running time.} Suppose $\mathtt{ApproxWidth}$ runs in time at most $m^{a}$ on any input of length $m$.
    The only non-trivial step of the above algorithm is that of computing the circuit $\tilde{C}$.
    Such an algebraic circuit has size at most $k \inparen{\size(C) + 2k \log (d+1)}$, using `repeated-squaring' gadgets for computing the appropriate powers of the $x_i$s.
    So $\size(\tilde{C}) \leq O_{\alpha}(\inparen{\sfrac{1}{\epsilon}}^2 \cdot \size(C))$, where the hidden constant depends on the approximation ratio $\alpha$ of $\mathtt{ApproxWidth}$.
    Next, the length of the unary representation for $(d+1)^k - 1$ has length $d^{O(k)} = d^{O_{\alpha}(\sfrac{1}{\epsilon})}$.
    Thus, step 3 runs $\mathtt{ApproxWidth}$ on an input of length at most $m_0^{\beta \cdot \inparen{\sfrac{1}{\epsilon}}}$, where $m_0$ is the length of the input $(C,n,d)$ and $\beta$ is a constant dependent on $\alpha$.
    The algorithm $\mathtt{WidthPTAS}$ thus runs in time at most $m_0^{a \cdot \beta \cdot \inparen{\sfrac{1}{\epsilon}}}$. This finishes the proof.
\end{proof}


\subsection*{Open questions}

We conclude with the following questions that we feel are natural to explore from here about the order-finding problem.

\begin{itemize}
    
    \item Does \autoref{alg:populate-graph} run in randomized polynomial time in the average case for fields of size $\poly(n,d,w)$?
    That is, is there a different argument that shows such a result?
    
    \item Are there efficient algorithms that approximate ROABP-width up to a $\poly(w,n,d)$ approximation factor?
    
\end{itemize}

\section*{Acknowledgements}

We thank the organizers of STOC 2022, where a subset of the authors first discussed this problem and made some preliminary progress.
VB also thanks Valentine Kabanets for their insightful discussions during a workshop on algebraic complexity in Dagstuhl, and he thanks the organizers of the workshop for facilitating this meeting.

\bibliographystyle{customurlbst/alphaurlpp}
\bibliography{masterbib/references,masterbib/crossref}

\appendix

\section{Definitions and Technical Statements}

\begin{definition}[Algebraic Circuit]\label{defn:algebraic-circuit}
  An \emph{algebraic circuit} is specified by a directed acyclic graph, with leaves (nodes with in-degree zero, called \emph{inputs}) labelled by field constants or variables, and internal nodes labelled by $+$ or $\times$.
  The nodes with out-degree zero are called the \emph{outputs} of the circuit. Computation proceeds in the natural way, where inductively each $+$ gate computes the sum of its children and each $\times$ gate computes the product of its children.

  The \emph{size} of the circuit is defined as the number of edges (or \emph{wires}) in the underlying graph.
\end{definition}

\end{document}